\documentclass[useAMS,plainnat,referee]{mn2e}

\usepackage{pdfpages}
\usepackage{graphicx}

\title[]{Infrared properties of Planetary Nebulae with [WR] and $wels$ central stars} 
\author[]{
C. Muthumariappan$^{1}$ \thanks{E-mail:muthu@iiap.res.in}
M. Parthasarathy$^{1,2}$
\\
$^{1}$Indian Institute of Astrophysics, Bangalore 560 034, India\\ 
$^{2}$National Astronomical Observatory of Japan (NAOJ), 2-21-1 Osawa, Mitaka,  Tokyo 181-8588, Japan
}

\begin{document} 

\date{Accepted ------. Received original form ------} \pubyear{2020}



\label{firstpage}
\pagerange{\pageref{firstpage}--\pageref{lastpage}}

\maketitle

   
\begin{abstract}

We report the IR properties of planetary nebulae with WR type and $wels$ central stars known to date and compare them with the IR properties of a sample of PNe with H-rich central stars. We use near-, mid- and far-IR photometric data from archives to derive the IR properties of PNe.  We have constructed IR colour-colour diagrams of PNe using measurements from $2MASS$, $IRAS$, $WISE$ and $Akari$ bands. [WR] PNe have a larger near-IR emission from the hot dust component and also show a tendency for stronger 12$\mu$m emission as compared to the other two groups. Cool AGB dust properties of all PNe are found to be similar. We derived the dust colour temperatures, dust masses, dust-to-gas mass ratios, IR luminosities and IR excess of PNe for these three groups. [WR] PNe and $wels$-PNe tend to have larger mean values for dust mass when compared to the third group. The average dust-to-gas mass ratio is found to be similar for the three groups of PNe. While there is a strong correlation of dust temperature and IR luminosity with the age for the three groups of PNe, the dust mass, dust-to-gas mass ratios and IR excess are found to be non-varying as the PNe evolve. [WR] PNe and $wels$-PNe show very similar distribution of excitation classes and also show similar distribution with Galactic latitude.

\end{abstract} 

\begin{keywords}
stars: AGB and post-AGB-- nebulae : planetary --- Stars : Circumstellar dust-- stars: evolution
\end{keywords}

\section{Introduction}
 
The planetary nebula phase (PN, plural PNe) occurs in the evolution of low- and intermediate mass stars after the end of AGB phase and before the resulting white-dwarf configuration. Based on the surface composition, the central stars of PNe (CSPNe) are divided into 
two groups i.e. hydrogen-rich and hydrogen-deficient (Mendez 1991). The central stars of most PNe are in the first group (Todt, Grafener \& Hamann 2006) which have thin hydrogen burning envelope and show hydrogen abundance close to the cosmic value on the surface
 (normal PNe, hereafter). About 30$\%$ of the CSPNe (Mendez 1991) belong to the second group where hydrogen is depleted in the stellar atmosphere and helium and carbon are the most abundant elements. From their spectral appearance one can distinguish between different types of hydrogen deficient CSPNe. A sizeable number of CSPNe (about 7$\%$) are known to exhibit Wolf-Rayet type spectrum (hereafter [WR] stars; see Crowther 2008). [WR] stars have hydrogen deficient atmospheres and display broad emission lines of highly ionized carbon, oxygen and helium. They exhibit fast stellar winds (with terminal velocities up to 3000 km s$^{-1}$)  with high mass-loss rates, up to two orders of magnitudes higher than the normal CSPN. Their spectra resemble the spectra of massive population I Wolf-Rayet stars of the carbon sequence. However, [WR] stars are low mass stars with degenerate structure. There is another group of CSPNe which show weak and narrow emission lines at the same wavelengths as those of [WR] stars which are called the weak emission lines stars ($wels$). However, unlike [WR] stars which are strictly hydrogen-deficient, the $wels$ do not necessarily have hydrogen-deficient atmospheres, though many of them do (Hajduk, Zijlstra \& Gesicki 2010). Weidmann, Mendez \& Gamen (2015) find that at least 26$\%$ of $wels$ are H-rich O stars. The PNe with [WR] stars ([WR] PNe) are stronger emitters in the $IRAS$ 12$\mu$m band (Zijlstra 2001) compared to other PNe and also show the presence of PAH emission (Szczerba et al. 2001). Some hydrogen deficient stars also show dominant absorption lines like PG 1159--035. 
 
The emission spectra of [WR] stars reflect physical processes which are different from other CSPNe. The WR-type emission lines of CSPN are interpreted as evidence of strong stellar winds. Detailed spectral analyses have shown that [WR] stars lose mass at a rate of 10$^{-6}$ M$_{\odot}$ yr$^{-1}$ (Leuenhagen, Hamann \& Jeffrey (1996); De Marco et al. 1997). This is up to two orders of magnitude higher than the typical mass-loss rate of a CSPN having hydrogen-rich atmosphere. A quantitative classification of [WR] stars based on the strengths of emission lines from ions with different ionization potential is related to the temperature evolution and can give the evolutionary connection between [WR] stars of different subtypes. Such quantitative classification schemes of [WR] stars currently in use were given by Crowther, De Marco \& Barlow (1998, CDB98) and by Acker \& Neiner (2003, AN03). CDB98 find a continuity in the [WR] classification scheme from [WC] to [WO] and suggest that the sequence is primarily due to excitation rather than abundance. All
subtypes are seen in a sequence of decreasing stellar temperature from [WO1] $\rightarrow$ [WO4] $\rightarrow$ [WC4] $\rightarrow$ [WC11]. AN03 conclude from this an evolution of  stars from post-AGB $\rightarrow$ [WC11] $\rightarrow$ [WC4] - [WO4] $\rightarrow$ [WO1]. Both CDB98 and AN03 find a sparsity of CSPNe in the [WC6-8] subclasses indicating a rapid evolution through this range or a different evolutionary path leading to two ionization sequences which is not yet understood. On the other hand, $wels$ do not have
such a classification scheme.

Though there are pronounced differences between the [WR] stars and the hydrogen rich CSPNe in their spectra and photospheric abundances etc., the PNe with [WR] stars have many properties which are very similar to those shown by the PNe with hydrogen-rich CSPNe. Particularly the similarities are seen in the elemental abundances, electron densities and temperatures and the morphologies of the PNe (Gorny \& Stansika 1995, Gorny 2001). This indicates that the [WR] phenomenon does not preferentially occur for massive progenitors. One difference between them could be the turbulence in the nebula: most [WR] stars (also some $wels$) have PNe with more turbulence which is almost absent for PNe with hydrogen-rich CSPNe as shown by Gesicki et al. (2006). They also show that the mass-averaged expansion velocities of [WR] PNe are higher than that for the PNe around $wels$ ($wels$-PNe) as well as the normal PNe. Higher expansion velocities for the [WR] PNe were also shown earlier by Tylenda \& Gorny (1993). 

The evolutionary path of low and intermediate mass stars which give birth to hydrogen-deficient [WR] stars and $wels$, and how and when it differs from the evolutionary path followed by hydrogen-rich CSPNe are not well understood. The observed differences of photospheric abundances of [WR] stars with the surface abundances of hydrogen-rich CSPNe show that [WR] stars have helium burning shells (Tylenda \&  Gorny 1993). [WR] stars are the observational testing for the He burning CSPN model. Hence, a process which can deplete hydrogen from the CSPNe and enrich their atmospheres with helium must play an important role. A process to deplete hydrogen in the stellar photosphere is the late or very late thermal pulse occurring after the star has left the AGB phase (born-again scenario; Werner \& Herwig 2006). The born-again ejecta show high fraction of carbonaceous dust, which requires large amount of carbon and very efficient dust formation (Borkowski et al. 1994, Zijlstra 2001). However, the abundances of born-again ejecta in the PNe A30 (Kingsburgh \& Barlow 1994) and A58 (Wesson et al. 2008) are oxygen and neon rich as expected from a nova-like ejection. Alternative theories were proposed on the origin of [WR] PNe, some of them were based on binary interaction (De Marco 2008). However, the close binary fraction of CSPNe observed for [WR] PNe could be much lower than that for the hydrogen-rich PNe  (Miszalski et al. 2009). In order to find if there is an evolutionary connection between [WR] stars and $wels$ and how they are different from the evolutionary path followed by the central stars of normal PNe, it is required to investigate in detail the gas and dust properties of PNe surrounding these stars. Parthasarathy et al (1998) found that the characteristics of the spectra of $wels$ are very similar to that of [WC]-PG1159 and they suggested an evolutionary sequence of [WCL] $\rightarrow$ [WCE] $\rightarrow$ $wels$ to PG 1159.

A comparative study between PNe with emission line central stars and PNe with hydrogen rich central stars was given earlier by Acker, Gorny \& Cuisinier (1996). Gesicky et al. (2006) have studied the kinematics and photo-ionization of nebula around [WR] stars. The chemical composition and other properties of 30 [WR] PNe and 18 $wels$-PNe were studied and were compared to a sample of normal PNe by Girard, Koppen \& Acker (2007). Gorny et al. (2001) analysed the IR properties of 49 [WR] PNe and addressed the dust content using near- and mid-IR photometric and spectroscopic observations gathered from the literature. More [WR] PNe and $wels$-PNe are known now and a few [WR] PNe have been reclassified. In this paper a uniform analysis of PNe surrounding [WR] and $wels$ central stars known to date is provided and compared with the IR properties of PNe around non-emission line central stars in order to address if their evolution is connected.  We have updated previous IR photometric studies with all [WR] PNe (99) known to date and derive the IR colour-colour diagrams of PNe and their nebular and dust properties. We also use $WISE$ and $Akari$ photometric observations in addition to $2MASS$ and $IRAS$ data. Far-IR $Akari$ data are used to trace the coolest dust of these PNe. The excitation classes of these PNe were also derived and are used. We make similar studies for a sample of  67 PNe with $wels$ central stars known to date and a sample of randomly selected 100 normal PNe. We constrain the properties of [WR] PNe, $wels$-PNe and normal PNe and compare them taking into account previous studies. Finally, we discuss the nature of these three groups of PNe and address if there are evolutionary connections between them. 

\section{Observational Data}

All the data required for this study were taken from the literature. The near-IR fluxes
 were calculated from the magnitudes at respective wavelengths which were obtained from the 
 $2MASS$ archive
(Cutri et al. 2003) archive. $IRAS$ (Neugebauer 1984) fluxes at 12-, 25-, 60- and 100$\mu$m 
and $WISE$ (Wright et al. 2010; Cutri et al. 2012) fluxes at 3.4-, 4.6-, 12- and 22 $\mu$m 
were used to span the mid- and far-IR emission. In addition, we have used archived  fluxes from $Akari$ (Ishihara et al. 2010)  at its 65-, 90-, 140- and 160-$\mu$m bands to trace the emission from cold dust 
down to $\sim$ 30K. The $2MASS$ and $Akari$ data used for this study are available at the NASA/IPAC Infrared Science Archive\footnote{http://www.irsa.ipac.caltech.edu/applications/BabyGator/}. All archival 
data we have used for this study have uncertainty measurements.
 
Our sample of [WR] PNe and $wels$-PNe are from the catalogue of spectral classification of CSPNe given by Weidmann $\&$ Gamen (2011). We have taken the nebular $H{\beta}$ fluxes and electron densities from Stasinska \& Szczerba (1999). Emission line fluxes of [OIII] and He II are obtained from Strasbourg-ESO Catalogue of Galactic Planetary Nebulae (Acker, Ochsenbein, Stenholm et al. 1992). Distances to PNe in our sample are from Frew, Parker \& Bojicic (2016; FPB16 hereafter) which were derived using $H_{\alpha}$ surface brightness of the nebula $vs$ its radius relationship. Recently, the distances for a sample of 16 PNe were given by Schonberner et al. (2018) using nebular angular expansion, and 128 PNe were given by Stanghellini et al. (2019) using GAIA parallaxes measurements (second GAIA release). These authors have also compared their values with the distances given by FPB16 and have shown that they are in reasonably good agreement, though for a few cases the values differ. Moreover, FPB16 have the distance measurements for all candidates in our sample. We have also taken the $E(B-V)$ values of PNe from FPB16. Some objects do not have IR data in some bands and/or they do not 
have the nebular $H{\beta}$ flux and/or density and/or distance measurements. 
 
\section{IR Colour-colour Diagrams} 

 IR colour-colour diagrams (CCDMs) are useful tools in studying the IR properties of celestial objects and are frequently being used 
by astronomers (Persi et al. 1987; Yerra et al. 2015). If the fluxes measured at two IR bands are $f_{\lambda1}$ and $f_{\lambda2}$ 
then the colour between the two respective bands with central wavelengths $\lambda1$ and $\lambda2$ can be defined using the relation 
given by Plets et al. (1997): 

\begin{equation}
 c_{\lambda1, \lambda2} = 2.5 \times log[(K_{\lambda1},z_{\lambda1}f_{\lambda2})/(K_{\lambda2},z_{\lambda2}f_{\lambda1})
\end{equation} 

 \noindent
 where, $K_{\lambda1}$ and $K_{\lambda2}$ are the colour correction factors for the bands with central wavelengths $\lambda1$ and 
 $\lambda2$ and $z_{1}$ and $z_{2}$ are their respective zero magnitude fluxes. For $IRAS$ and $Akari$ bands, where the measurements
 are given in fluxes, we have calculated the colours using this relation. For $2MASS$ and $WISE$ bands, where the measurements are 
 in magnitudes, we have taken the difference of magnitudes at the two bands as the colour. The colour correction factors for
 $IRAS$ photometry were taken from Neugebauer et al. (1984) corresponding to a power law spectral index of -1 and their zero magnitude 
 fluxes are 28.3-, 6.73-, 1.19- and 0.43 Jy respectively at 12- 25-, 60- and 100$\mu$m bands (taken from $\it{http://www.ipac.caltech.edu/IRASdocs/exp.sub.ch6/C2a.html}$). For the $Akari$ bands, the colour correction factors for spectral index -1 at 65-, 90-, 140- and 160 
 $\mu$m  and the zero magnitude fluxes at these bands were taken from Shirahata et al. (2009). We have calculated the colours between 
 two $IRAS$ bands if the fluxes at these bands are given with uncertainty. The maximum error in the $IRAS$ fluxes of our
 sample is 16$\%$. Though there are many PNe which have $WISE$ as well as $IRAS$ measurements, some have either of that. In the following
 sections we discuss the CCDMs of near-IR, mid-IR and in far-IR bands for [WR] PNe, $wels$-PNe and normal PNe. 
 
\subsection{Near-IR colour analysis}

  Near-IR radiation from PNe consists of atomic emission lines and free-free and free-bound nebular continuum. In
  addition to this, hot dust present in the nebula also emits near-IR continuum and near-IR CCDM can bring out the dominant source of radiation at this wave band. We used the $2MASS$ photometric measurements to calculate near-IR colours. Though  
  some PNe in the sample have near-IR data from other sources, we have taken the 2MASS data to be consistent and to 
  make the study uniform with other PNe which have only 2MASS data. The emission from PNe at this band can suffer 
  significant extinction by interstellar dust on its way to the observer. Hence, interstellar reddening corrections 
  were applied to the photometric measurements using the relations given by Whitelock (1985). The $E(B-V)$ values 
  required for this correction were obtained from FPB16. We have made near-IR colour 
  analysis for 59 [WR] PNe, 42 $wels$-PNe and 63 normal PNe for which $2MASS$ photometric measurements are available 
  in the literature.
 
  Fig 1 shows the extinction corrected colour-colour diagram of PNe plotted between $[J-H]$ and $[H-K]$ ($[J-H]_{0}$ 
  and $[H-K]_{0}$) along with their errors. The errors in near-IR colours were calculated from the photometric 
  errors in their respective bands obtained from archive and the error in $E(B-V)$ from FPB16. 
  For reference we have also drawn a line in Fig 1 showing the stellar 
  photospheric colours of main sequence stars (Koorneef 1983; S region); the bottom of the line represents the
  B-type stars and the top represents the M-type stars. Also shown in the figure is the nebular box  (N
  region) which corresponds to the near-IR emission in the form of free-free, free-bound and ionic emission lines
  from the gaseous nebula (Whitelock 1985). A dashed line seen in Fig 1 represents the dust colours due to dust 
  continuum emission (D region). This was computed with an emissivity exponent of 1 ($F_{\nu} = \nu B_{\nu}(T)$), 
  which better represents the wavelength dependence continuum emission for circumstellar dust (Stasinska \& Szczerba 1999;
  Muthumariappan, Kwok \& Volk 2006). The bottom of this line corresponds to a dust colour temperature of 2000K and 
  the temperature decreases along the line to 1000K at the top. Hence the location of a PN in this figure allows us 
  to see the relative contribution from its different components (N, S or D) to the near-IR radiation.\\
  
  \noindent
  Gorny et al. (2001) found from their near-IR CCDM that 9 [WR] PNe, which comprise about 30$\%$ of their [WR] PNe 
  sample are located near the stellar main sequence line. They suggested that the  similar colours to the main 
  sequence stars imply that these PNe could have companion(s) to their [WR] central stars. It is also possible 
  that the main sequence star may be a neighbouring star to the PN contaminating the near-IR flux. In our 
  sample there are 8 [WR] PNe (PM 1-89 is not included here as it is identified as [WR] with a peculiar spectrum by 
  Weidmann \& Gamen 2011) which makes $\sim$14$\%$ of our [WR] PNe sample. However, there are 12 candidates in the 'NS' 
  region (see Fig 1a) where the contribution from both nebular and stellar components are important. Hence the
  [WR] PNe which have significant contribution from the main sequence stars becomes $\sim$34$\%$ in our sample and 
  is $\sim$43$\%$ in the sample considered by Gorny et al. (2001). About 40$\%$ of [WR] PNe in the sample of Gorny et al.
  (2001) fall inside or around their nebular box, where the emission line of helium triplet at 1.083$\mu$m can be a 
  major contributor (Whitelock 1985). However, our sample has only about 17$\%$ of [WR] PNe which are located inside or 
  near to the nebular box. Hence most [WR] PNe are distributed away from the nebular box. In Fig 1a $\sim$42$\%$ of 
  [WR] PNe are located in the 'ND' region, where significant emission come from both the hot dust 
  component and the nebular component. Fig 1b shows the near-IR CCDM for the subtypes of [WR] PNe, whose central stars
  were identified as late type, early type or intermediate type [WR] stars. The figure shows that there is no
  significant difference in the distribution of PNe with [WCL] and [WCE] stars in the 'NS' and 'ND' regions.  Fig 1b
  also shows that none of the PNe with [WCL] stars are located inside the 'N' box, however, a few PNe with [WCE] stars
  are located inside 'N' box and some are seen closer to it. This could imply that more evolved [WR] PNe show
  brighter nebular emission. \\

  As it is shown in Fig 1a, $wels$-PNe are concentrated in or close to the nebular box, in contrast to the [WR] PNe. Out 
  of 42 $wels$-PNe, only 3 ($\sim$7$\%$) are located near the stellar line and 5 ($\sim$12$\%$) of them are seen in the 
  'NS' region. Their concentration in the 'ND' region is, however, significant (13, which is $\sim$ 31$\%$). The near-IR
  colours of normal PNe are mostly distributed near the stellar line, inside  or closer to the nebular box and in the 
  'NS' region. There are only a few of them located in the 'ND' region (10 out of 63 normal PNe in our sample; that is 
  $\sim$16$\%$). Those PNe falling close to the stellar line or in the 'NS' region of Fig 1a have the possibility of
  having binary nuclei, as suggested by Gorny et al (2001),  which needs to be investigated. 
   
\begin{figure*}  
\includegraphics[width=\textwidth,keepaspectratio]{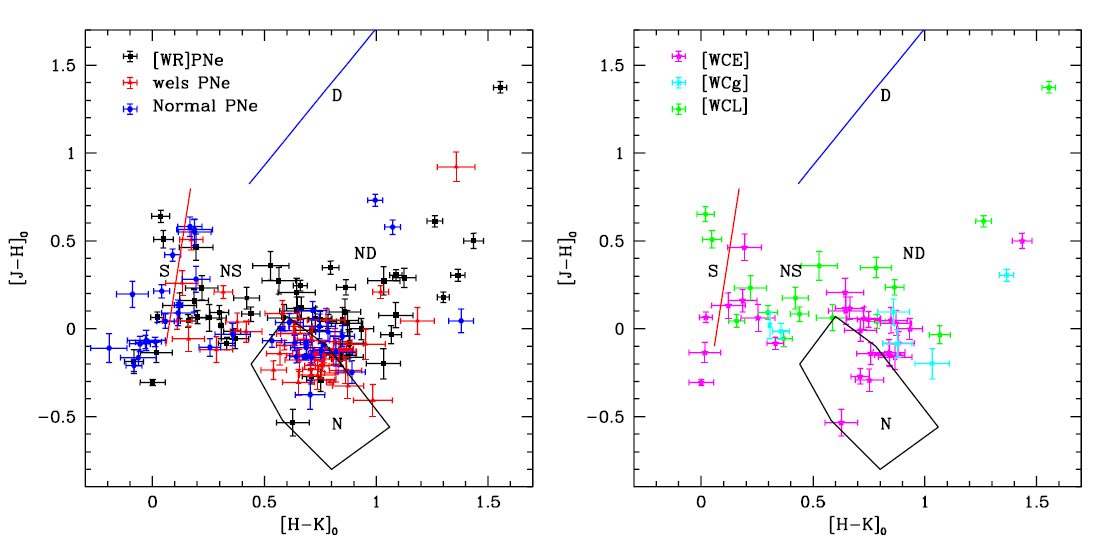}
\caption{Extinction corrected near-IR colour-colour diagram of PNe using $2MASS$ magnitudes. a) $Left$: for [WR] PNe, $wels$-PNe and normal PNe. $Right$: PNe with [WCL], [WCE] and intermediate type [WCg] central stars. The solid line 'S' in red is the stellar line, broken line 'D' in blue is the dust line and 'N' is the nebular box. See the text for details.}  
\end{figure*}


\begin{table}
\label{tbl2}
\vspace{0cm}
\caption{Mean values of colours between different IR bands and their dispersion for [WR] PNe, $wels$-PNe and normal PNe. The number of PNe in each class for which the colours were estimated are also shown.} 
\vspace{0cm}
\begin{tabular}{lrrr}
\hline
\hline
 Colour   & [WR] PNe & $wels$-PNe   & normal-PNe    \\
          &  (No. of PNe)       &  (No. of PNe)   &  (No. of PNe)     \\
\hline
 $WISE$   &          &         &        \\
$[3.4-4.2]$   &   0.88$\pm$0.58 (63) & 0.81$\pm$0.29 (47)& 0.77$\pm$0.55 (72) \\
$[4.2-12]$   &  5.17$\pm$1.20 (63)& 5.29$\pm$0.80 (47) & 4.57$\pm$1.50 (71)  \\
$[12-22]$   &  3.37$\pm$0.90(73) & 3.65$\pm$0.55 (55) & 3.65$\pm$1.10 (80) \\
 $IRAS$   &          &         &          \\
$[12-25]$   &  3.45$\pm$0.63 (55)  & 3.95$\pm$0.50 (27) & 3.73$\pm$0.70 (41)  \\ 
$[25-60]$   &  2.41$\pm$0.74 (78) & 2.38$\pm$0.54 (52) & 2.38$\pm$0.86 (67) \\
$[60-100]$  &  0.21$\pm$0.50 (29) & 0.17$\pm$0.65 (18) & 0.68$\pm$0.54 (43) \\
$Akari$   &          &         &             \\
$[65-90]$   &   0.35$\pm$0.37 (50) & 0.38$\pm$0.22 (35) & 0.70$\pm$0.50 (53)  \\
$[90-140]$  &   0.07$\pm$0.70 (45) & 0.34$\pm$0.62 (26)  & 0.55$\pm$0.79 (48)  \\
$[140-160]$  &  -0.11$\pm$1.10 (31)  &-0.18$\pm$0.90 (19)  & -0.46$\pm$1.20 (41) \\
          &         &         &             \\
\hline
\end{tabular}
\end{table} 

\vspace{2.0cm} \protect \begin{figure*} 
\includegraphics[width=\textwidth,keepaspectratio]{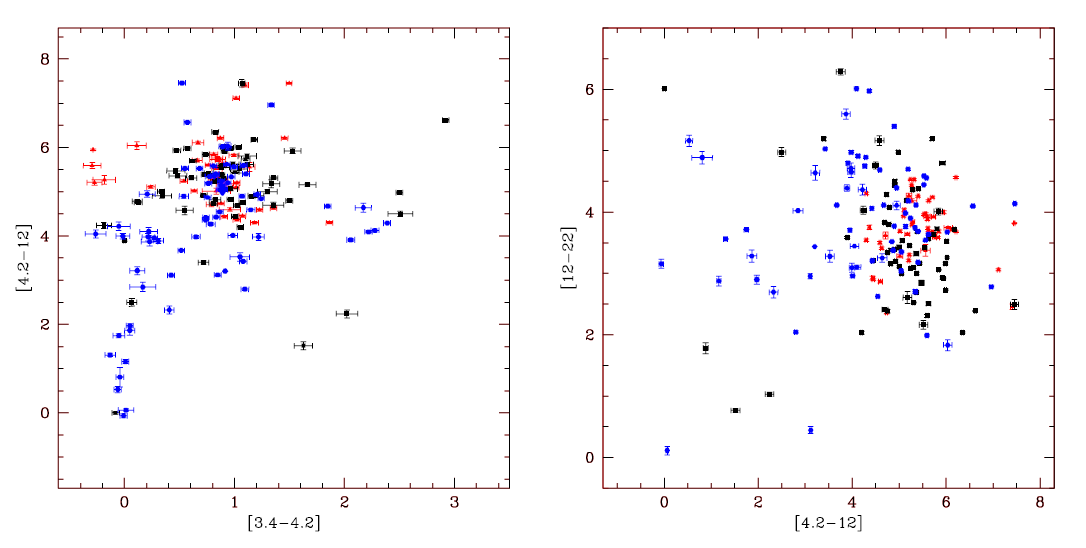}
\caption{$WISE$ colour-colour diagrams of PNe. Data description as given in Fig. 1. See the text for details.} 
\end{figure*} 
 
\vspace{2.0cm} \protect 

\begin{figure*} 
\includegraphics[width=\textwidth]{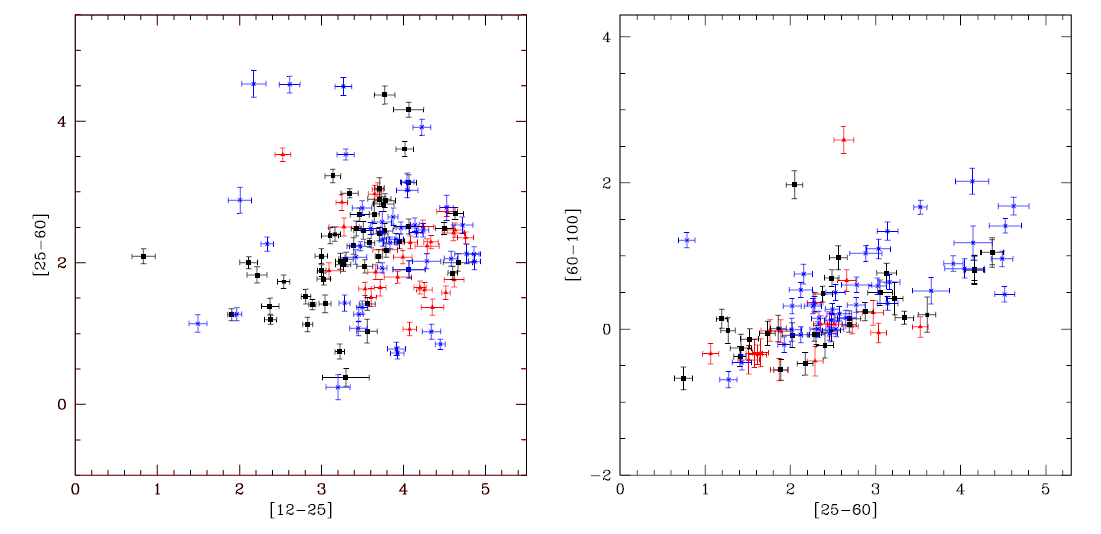}
\caption{ a (left) and b (right): $IRAS$ colour-colour diagrams of [WR] PNe, $wels$-PNe and normal-PNe; data description as in Fig 1. See the text for details.} 
\end{figure*} 

\subsection{Mid-IR colour analysis}

 Mid-IR radiation of PNe originate mostly from their warm, thermal equilibrium dust component which are at a typical
 colour temperature of $\sim$ 150K - 300K. This dust was created in the cool and dense atmosphere of the AGB progenitor and 
 is being heated by the hard UV radiation from the central star of the PN. The warm dust emission of a PN can be better 
 traced by the N band (10$\mu$m) flux and also is seen in the Q band (20$\mu$m). In addition to this, the N band also 
 includes the emission features of the Polycyclic Aromatic Hydrocarbons (PAH) and the atomic forbidden transitions.
 We construct the CCDMs using $WISE$ bands at 3.6-, 4.5-, 12- and 22$\mu$m and $IRAS$ bands at 12-, 25$\mu$m to
 study the properties of hot and warm AGB dust. The 3.6$\mu$m and 4.5$\mu$m $WISE$ bands respectively 
 cover the PAH features at 3.3$\mu$m and 4.6$\mu$m and also have significant continuum emission from the hot dust
 component. The PAH features covered by the $WISE$ 12$\mu$m band are at 7.7-, 8.6- and 11.3$\mu$m and by the $IRAS$ 
 12$\mu$m band are at 8.6- and 11.3$\mu$m. In addition to the PAH features, the 12$\mu$m bands can also have atomic
 line fluxes of [Ar V] 7.9$\mu$m and [Ar IV] 9.0$\mu$m, [S IV] 10.5$\mu$m, [Ne II] 12.8$\mu$m and [Ne V] 14.3$\mu$m. 
 Relatively the 25$\mu$m band of $IRAS$ and the 22$\mu$m band of $WISE$ are least contaminated by atomic emission 
 lines. Mid-IR CCDMs of PNe between these bands can be used to trace the relative emission contribution of the warm
 dust continuum, the PAH and atomic emissions in PNe. This can also bring out the differences in the emission components 
 between [WR] PNe, $wels$-PNe and the normal-PNe. $WISE$ colours are computed for these three groups of PNe, and are 
 shown in Fig 2. Errors in the mid-IR colours are calculated from the errors in photometric measurements at their 
 respective bands obtained from archive. Fig 3a shows the CCDMs of the three groups of PNe in the  $IRAS$ 
 mid-IR bands. Table 4 shows the average values of mid-IR colours with their dispersions (or the range) of [WR] PNe, 
 $wels$-PNe and normal-PNe. \\    

\noindent  
 Based on these colours we discuss the similarities and differences in the properties of warm AGB dust
 between these three groups of PNe. The mean values for $WISE$ [3.3-4.6] and [4.6-12] colours of [WR] PNe 
 are similar to those values derived for the other two groups of PNe, taking their dispersion into account.
 The mean values of $IRAS$ [12-25] and $WISE$ [12-22] colours of $wels$-PNe are similar to their respective
 values of normal-PNe, however, [WR]PNe show somewhat lower values (see Table 1). Though the dispersions of 
 mean values in $WISE$ [12-22] colours are large, $IRAS$ [12-25] colours show relatively smaller dispersions 
 and indicates that [WR]PNe have a tendency towards a lower mean value than non-[WR] PNe. This also in turn 
 indicates a possibility of a brighter 12$\mu$m band for [WR] PNe as compared to the other two groups. This 
 was also suggested earlier by Zijlstra (2001) for a set of 40 [WR] PNe using $IRAS$ data. Though the 12$\mu$m
 band is contaminated by atomic lines, the presence of strong PAH emission in this band can dominate over the 
 atomic line fluxes, as discussed by Stasinska \& Szczerba (1999) the emission from atomic lines become more 
 important only for more evolved PNe. If we take that the contribution from atomic lines are not very important
 or that they are similar for all the three groups considered in this study, then the tendency for a brighter 
 12$\mu$m band shows that [WR] PNe likely to have stronger PAH emission as compared to $wels$-PNe and normal PNe. 
 All the three CCDMs show that $wels$-PNe are least scattered in the mid-IR colour-colour plane. The 
 $wels$-PNe are distributed with a spread which is $\sim$ 50$\%$ lower than that for the normal-PNe, and 
 the [WR] PNe are distributed similar to the normal PNe (see Fig 2 and Fig 3a). It shows that the mid-IR
 colours of $wels$-PNe group do not evolve that significantly as compared to that of the other two groups. 
 
\subsection{Far-IR colour analysis}

 The continuum emission from cool AGB dust in PNe (T$_{d}$ less than 150K) can be best traced by the 25- and
 60$\mu$m bands of $IRAS$, where the contamination due to dust features and atomic emission lines are minimal
 (Stasinska \& Szczerba 1999). Also the dust continuum emission of PNe peaks in this wavelength domain. A
 study of this cool dust in [WR] PNe, $wels$-PNe and normal PNe will bring out the differences in their AGB
 evolution, if any. The mean values and dispersions of [25-60] colour for these three groups of PNe are
 calculated and are shown in Table 1 Also Fig 3a shows a plot of $IRAS$ [12-25] $vs$ [25-60] colours for
 them along with their errors. As can be seen  in Table 1, the mean values of [25-60] colour for the three
 groups of PNe are nearly the same. The similarity in the [25-60] colour for [WR] PNe and non-[WR] PNe was
 also reported earlier by Zijlstra (2001). Our study hence supports the finding that the AGB dust properties
 inferred by [25-60] colours are similar for [WR] PNe, $wels$-PNe and normal PNe. Far-IR colours of the
 coolest dust in PNe are studied using $IRAS$ band at 100$\mu$m and $Akari$ bands. Fig 3b shows the $IRAS$
 [25-60] $vs$ [60-100] colours and $Akari$ colours are plotted in Fig 4 along with their errors. As can be
 seen in Table 1, the mean values of $IRAS$ [60-100] colour and $Akari$ [65-90] colours are similar for [WR]
 PNe and $wels$-PNe taking their dispersion into account, but normal PNe have a tendency towards higher value
 (Table 1). Mean values of $Akari$ [90-140] and [140-160] colours are also shown in Table 1.

In Fig 5 we show the Galactic distribution of [WR] PNe, $wels$-PNe and normal PNe at different latitudes. 
 To see if their distributions are similar, we have performed a two sample Kolmogorov-Smirnov test between
 these groups of PNe. The value of the test statistics between the distributions of [WR] PNe and $wels$-PNe
 is $D=0.017$ and the probability to find the test statistics below the critical value is $P=0.99$. This
 shows that these two distributions are similar. However, the test performed between the distributions of
 [WR] PNe and normal PNe shows $D=0.138, P=0.63$ and between the distributions of $wels$-PNe against normal
 PNe shows the values of $D=0.127, P=0.58$. This brings out that the distribution of normal-PNe with Galactic
 latitude is somewhat different from that of [WR] PNe and of $wels$-PNe. [WR] PNe and $wels$-PNe are
 concentrated near the Galactic plane, however, normal PNe are commonly seen at higher latitudes. This 
 is in contrast with the suggestion given by Acker, Gorny \& Cuisinier (1996). 
  
  
\begin{table*}
\begin{minipage}{120mm}
\label{tbl1}
\caption{Nebular parameters derived for [WR] PNe (see the text for details)}
\vspace{1cm}
\begin{tabular}{@{}lrrrrrrrr}
\hline 
\hline 
PNG & class    & $T_{d}$ & $M_{d}$ & $m_{d}/m_{g}$ & $ L_{IR}$ & IRE & Exc. \\
    &          & (K)     & (10$^{-4}$M$_{\sun}$)& (10$^{-3}$) & (L$_{\sun}$)&   & class \\
\hline
000.4-01.9 & [WC5-6] & 85 & 1.52$\pm$0.73 & 4.20$\pm$0.73 & 535$\pm$251 & 1.55$\pm$0.22 & 2 &\\
001.2-05.6 & [WO1-2] & 76 & 1.00$\pm$0.41 & --  &  201$\pm$80& -- & - & \\
001.5-06.7 & [WC9]pec & 170 & 0.52$\pm$0.32 & -- & 5825$\pm$2830& 15.15$\pm$2.12 & 8  \\
001.7-04.4 & [WC11]   &  73 & 5.42$\pm$2.47 & -- & 890$\pm$436 & -- & 1 &\\
002.2-09.4 & [WO4]pec &  92 & 1.75$\pm$0.73 & 6.02$\pm$1.0 & 914$\pm$368 & 3.75$\pm$0.53 &1 &\\
002.4-03.7 & [WC11]   & 105 & 1.38$\pm$0.62 &11.1$\pm$2.1 & 1392$\pm$612& 5.23$\pm$0.74 & -& \\
002.4+05.8 & [WO3]    &  85 & 0.85$\pm$0.37 &3.12$\pm$0.6 & 300$\pm$131 & 1.23$\pm$0.17 &5 &\\
002.6+05.5$^{1}$& [WC4]    &  --  & -- &  --   &  --    & -- &11& \\
003.1+02.9 & [WO3]    &  81 & 1.89$\pm$0.82&4.95$\pm$0.86  & 522$\pm$221 & 1.73$\pm$0.24 &7 & \\
004.8-22.7 & [WC4]    &  115& 1.84$\pm$0.75  & -- &2934$\pm$1250 & --   &- & \\
          &               &                   &     &       &        &     &   & \\
004.9+04.9 & [WC5-6]  &  90 & 2.13$\pm$1.0 &6.0$\pm$1.0 & 997$\pm$478 & 2.53$\pm$0.36 &2 & \\
005.9-02.6 & [WC8]    &  -- &  --   & -- &--  -- & --   & -& \\
006.0-03.6 & [WC4]    &  86 & 1.73$\pm$0.72  & 2.9$\pm$0.5&642$\pm$340 & 1.63$\pm$0.23 &3 & \\
006.8+04.1 & [WC4]    &  89 & 2.01$\pm$0.89  &2.6$\pm$0.45 &888$\pm$385 & 1.67$\pm$0.24 & 6& \\
007.8-03.7 & [WC]     &  -- &  --   & -- & --     & --  &9 & \\
009.8-04.6 & [WO2]    &  84 &  0.61$\pm$0.39 & 1.1$\pm$0.22 &202$\pm$110& 2.69$\pm$0.70 &10& \\
011.1-07.9 & [WC12]   &  99 & 1.40$\pm$0.62& --  & -- &1055$\pm$447 & -- &  \\
011.9+04.2 & [WO4]pec &  94 & 1.60$\pm$0.57& 2.13$\pm$0.5 &929$\pm$418 & 5.52$\pm$0.75 &-& \\
012.2+04.9 & [WC10]   &  99 & 4.30$\pm$1.9 & -- &3239$\pm$1458&-- & - & \\
016.8-01.7 & [WR]     &  77 &  --   &-- & --    & -- & - & \\      
          &           &                   &     &       &        &     &   & \\
017.9-04.8 & [WO2]    &  77 & 0.67$\pm$0.35 &8.7$\pm$1.5 & 145$\pm$64 &1.74$\pm$0.25& 11 &\\
019.7-04.5 &  [WC4]   &  87 & 3.41$\pm$0.55& 4.9$\pm$0.6 & 1348$\pm$741 & 2.80$\pm$0.74 & 3& \\
020.9-01.1 & [WO4]pec &  90 & 1.33$\pm$0.63 & 2.5$\pm$0.47& 625$\pm$287 & 1.80$\pm$0.25 &2& \\
021.0-04.1 & [WC4]    &  --  &   --& -- &   --   & -- & - & \\
027.6+04.2 & [WC7-8]  & 140 & 2.23$\pm$1.1& 3.2$\pm$0.55 & 9499$\pm$4788& 7.50$\pm$1.06 & 2& \\
029.0+00.4 & WC       &   -- &   -- & -- &   --   & --& 1 & \\
029.2-05.9  & [WO4]   &  91 & 1.1$\pm$0.47 &5.7$\pm$0.98 & 547$\pm$226 & 6.8$\pm$0.96 &3 &\\
037.5-05.1 & [WCE]    &  89 & 10.80$\pm$5.3 & -- & 4782$\pm$2530& -- &- & \\
037.7-06.0 & [WC4] &  --  &  --   & -- & --    & --& - & \\
048.7+01.9  &[WC4]   &  85 & 0.91$\pm$0.45 & 9.18$\pm$1.54 & 321$\pm$143 &3.11$\pm$0.44& 2& \\
           &                &                  &     &       &        &     &  & \\
060.4+01.5 & [WC11]   & 110 & 8.96$\pm$3.9 & -- &  --    & -- & - & \\
061.4-09.5 & [WO2]    & 89  & 0.29$\pm$0.11 &0.94$\pm$0.15 & 126$\pm$51 & 2.98$\pm$0.42& 11& \\
064.7+05.0 & [WC9]    & 111 & 3.66$\pm$1.5 & 5.4$\pm$0.9& 4884$\pm$1960& 6.75$\pm$0.95&1& \\ 
068.3-02.7 & [WC9]    &  97 & 13.0$\pm$8.5 & 8.8$\pm$1.4& 8835$\pm$5770& 14.5$\pm$2.05& - \\
089.0+00.3 & [WO3]    &  78 & 1.47$\pm$0.63 &4.75$\pm$0.94 & 336$\pm$137 & 2.00$\pm$0.28 & 9& \\
089.8-05.1 & [WR]     & 124 & 2.08$\pm$0.95 & 5.9$\pm$1.1& 4816$\pm$2127& 5.22$\pm$0.74 &9 &\\
093.9-00.1 & [WC11]   & 132 & 2.11$\pm$1.12 & -- & 6687$\pm$3536& --&- & \\
094.0+27.4 & [WO]     &  --  & --  & -- &  --    & -- &12+& \\
096.3+02.3 & [WC4-6]  &  80 & 1.39$\pm$0.6 &2.6$\pm$0.42 & 361$\pm$153 & 2.15$\pm$0.3& 10& \\
118.0-08.6 & [WC]     &  89 & 0.76$\pm$0.31 &3.9$\pm$0.63 & 335$\pm$138 &1.60$\pm$0.23 & 5 & \\
           &               &                   &     &       &        &      &  & \\ 
118.8-74.7 & WC OVI   &  70 & 0.10$\pm$0.05 & -- & 13$\pm$7.0  & -- &12& \\
120.0+09.8 &[WC8]     & 101 & 0.68$\pm$0.28  & 1.1$\pm$0.2& 564$\pm$226 & 3.74$\pm$0.53 &- & \\
130.2+01.3 & [WO4]    &  79 & 0.92$\pm$0.41 &3.4$\pm$0.52 & 223$\pm$102 & 1.40$\pm$0.2 &7 &\\
144.5+06.5 & [WO4]    &  75 & 0.32$\pm$0.15 & 1.53$\pm$0.25& 60$\pm$26 & 1.17$\pm$0.16 &3 &\\
146.7+07.6 & [WC11]   & 136 & 0.68$\pm$0.28 &0.78$\pm$0.12 & 2516$\pm$1042& 9.68$\pm$1.37 &-&\\
189.1+19.8 & [WO1]    &  87 & 0.39$\pm$0.17 & 1.1$\pm$0.2& 155$\pm$64 & 2.70$\pm$0.39 &-& \\
189.8+07.7 & WC-OVI   &  63 & 5.59$\pm$2.4 &18.3$\pm$3.6 & 439$\pm$184 & 3.67$\pm$0.52&8& \\
216.0+07.4 & [WC4]   &  62 & 1.10$\pm$0.48 &-- &  80$\pm$34 & -- & - &\\
222.8-04.2 & [WC7-8] & 103 & 0.35$\pm$0.19  &-- &  321$\pm$178& -- & - & \\
243.3-01.0 & [WO1]    &  90 & 0.58$\pm$0.23 &2.6$\pm$0.4 & 272$\pm$109 & 3.2$\pm$0.45&10& \\
           &               &                   &     &       &        &      &  & \\ 
\hline
\end{tabular}
\end{minipage}
\end{table*}

\begin{table*}
\begin{minipage}{120mm}
\label{tbl2}
\vspace{1cm}
\begin{tabular}{@{}lrrrrrrrr}
\hline
\hline
PNG & class    & $T_{d}$ & $M_{d}$ & $m_{d}/m_{g}$ & $ L_{IR}$ & IRE & Exc. \\
    &          & (K)     & (10$^{-4}$M$_{\sun}$)& (10$^{-3}$) & (L$_{\sun}$)&   & class \\
\hline
270.1-02.9 & [WC5-7] & 101 & 0.69$\pm$0.28 & 1.2$\pm$0.21& 565$\pm$226 & 3.75$\pm$0.53 & - & \\
272.8+01.0 & [WC9-10] &  --  &  --   & -- & --     & --&- \\
278.1-05.9 & [WO2]    &  94 & 0.60$\pm$0.24  & 0.91$\pm$0.14& 347$\pm$139 & 1.54$\pm$0.22 & 8 &\\
278.8+04.9 & [WO1]    & 104 & 0.81$\pm$0.35 &2.45$\pm$0.4 & 774$\pm$329 & 6.85$\pm$0.97&- \\
285.4+01.5 & [WO4]    & 114 & 1.49$\pm$0.55  & 3.17$\pm$0.51 & 2270$\pm$1250& 5.08$\pm$0.70& 3 &\\
284.2-05.3 & [WO3] &  -- &  --   & -- & --   & -- & - & \\
286.3+02.8 & [WO3]    &  97 & 0.22$\pm$0.11 &0.71$\pm$0.12 & 149$\pm$72 & 6.94$\pm$0.98&12& \\
291.3+08.4 & [WO4]-[WC4]    &  73 &-- & 7.08$\pm$1.41  & 1161.21& --& - \\
291.3-26.2 & [WC10]   &   -- &  -- & -- & --     & --&- \\
292.4+04.1 & [WC5-6]  &  85 & 2.53$\pm$1.0 & 8.7$\pm$1.4& 887$\pm$358 & 5.19$\pm$0.73 &2  \\
          &               &                   &     &       &        &     &   &  \\
297.0+06.5 & [WO2] &  --  &  --   & -- &   --   & -- & - &  \\
300.7-02.0 & [WC4] & 104 & 2.66$\pm$1.2 & 4.15$\pm$0.56 & 2558$\pm$1106 & 2.72$\pm$0.45 & 4 & \\
302.0-01.6 & [WC9] &  --  &  --   & -- & -- &  --   & -- &  \\
306.4-00.6 & [W03]pec &  84 & 0.34$\pm$0.15 &1.06$\pm$0.2 & 114$\pm$48 & 1.78$\pm$0.25 &- & \\
307.2-03.4 & [WO1]    &  77 & 0.26$\pm$0.1 & 0.41$\pm$0.07 & 57$\pm$22  & 1.21$\pm$0.17 &10& \\
308.5+02.5 & [WC4]    &  --  &   -- &-- & --     & -- & - & \\
309.0-04.2 $^{2}$&[WC9]& 91 & 0.88$\pm$0.36 &3.7$\pm$0.62 & 435$\pm$176 & 8.04$\pm$1.13 &3 & \\
309.1-04.3 & [WO4]    &  94 & 2.44$\pm$1.0 &14$\pm$2.4 & 1422$\pm$584& 3.53$\pm$0.5 &3& \\
309.8-01.6 & [WC4] &  --  &  --   &--&   --   & --& - &  \\
313.9+02.8 & [WC9]    &  82 &  -- &--&   --   & -- & - & \\ 
          &               &                   &     &       &        &     &   & \\
319.6+15.7 & [WR]     &  63 & 2.2$\pm$0.92 &3.7$\pm$0.68& 135$\pm$54 & 1.47$\pm$0.21 &5 & \\
321.0+03.9 & [WC10]   & 120 & 39.0$\pm$16 &44.8$\pm$7.7 & 9700$\pm$5300 & $>$25.00 &1& \\
324.0+03.5 & [WO4]pec &  --  &  --   & -- &  --    & -- &- & \\
327.5-02.2 & [WR]     &  --  & --    & -- &   --   & -- &- & \\ 
327.1-02.2 & [WC9]    & 126 & 1.11$\pm$0.53 &19.6$\pm$3.5  & 2767$\pm$1298& 6.83$\pm$0.97& 1& \\
332.9-09.9 & [WC10]   & 107 & -- &  --    & -- &--& - \\
336.2-06.9 & [WO4]    & --   &  -- & --  &  --    & --&3 & \\
336.5+05.5 & [WO4] &  --  &  --  & -- &   --   & --& - & \\
337.4+01.6 & [WC9]    &  99 &  -- & --  &  --    & -- & 1 &\\
341.5+12.1 & [WC3]    &  63 &  -- & --  &  --    & -- &-& \\
          &               &                   &     &       &        &     &   & \\
348.4+04.9 &[WC10] &  108&  0.37$\pm$0.16 &--& 431$\pm$176& -- & - & \\
350.1-03.9 & [WC4-5]  &  82 & 0.63$\pm$0.29&25.4$\pm$4.4  & 184$\pm$85& 7.01$\pm$0.98& 9& \\
350.9+04.4 & [WC11]   & 124 & 0.75$\pm$0.32 &1.1$\pm$0.2& 1728$\pm$737& 4.40$\pm$0.62 & 1& \\
351.5-06.5 & [WO2]    &  73 & 3.23$\pm$1.3 &--  & 530$\pm$219 & -- &- & \\
352.9+11.4 & [WC11]   & 106 & 5.3$\pm$2.2 &--   &  5618$\pm$2336 & --&-& \\
355.4-04.0 & [WO2]  &  89 &  0.47$\pm$0.2 &1.9$\pm$0.3 & 206$\pm$87 & 7.30$\pm$1.0 &11& \\
355.2-02.5 & [WC4]    &  --  &  --  & -- &   --   &--& 3& \\
355.9+03.6 & [WC11]   & 215 & 0.11$\pm$0.06& --  & 4165$\pm$1936& -- & - & \\
355.9-04.4 & [WO3]  &   -- &   --  & --&   --   & --& - & \\
356.0-04.2 & [WO2-3] &   -- &   -- & -- &   --   & --& - & \\
          &               &                   &     &       &        &     &   \\
356.1+02.7 & [WC5-6] & 146 & 0.92$\pm$0.5 & --  &  4811$\pm$2604& --& - & \\
356.8-03.6 & [WC11]   &   -- &   --  &-- &  --   & --& - & \\
357.1-04.7 & [WC11]   & 102 & 7.20$\pm$3.4 &21$\pm$4  & 6263$\pm$2926& 17.37$\pm$2.45 &1 & \\
357.3+03.3 & [WC11]   &  92 & 0.78$\pm$0.33 &-- & 405$\pm$168 & -- & - & \\
357.4-03.2 & [WO2-3] &  --  &   -- &--  &  --    & -- & 7&  \\
358.3-21.6 & [WO3]    &  81 & 0.89$\pm$0.37&2.3$\pm$0.4 & 245$\pm$100& 1.36$\pm$0.19 & 9& \\
359.8+03.5 & [WC4]  &   -- &   --  &  -- & --   & -- & - & \\
359.9-04.5 & [WC4] &  99 & 1.66$\pm$0.61  &2.42$\pm$0.41 & 1247$\pm$730& 2.85$\pm$0.42 &3 &\\
359.8+05.6$^{1}$ & [WC11]   & 104 & 1.15$\pm$0.5 &6.7$\pm$1.2  & 1109$\pm$472& 6.08$\pm$0.85 &1 & \\ 

\hline
\end{tabular}
\\
{\hspace{0cm} \sevensize $^{1}$ Escudero \& Costa (2001);
\sevensize $^{2}$ Garcia-Rojas, Pena, Morriset et al. (2012);
\sevensize $^{3}$ Phillips (2005);
\sevensize $^{4}$ Escudero, Costa \& Macial (2004)\\
\sevensize $^{5}$ Balick, Perinotto, Maccioni et al. (1994);
\sevensize $^{6}$ Bohigas (2003);
\sevensize $^{7}$ Kwitter \& Henry (2001);
\sevensize $^{8}$ Madsen, Frew \& Parker (2006);\\
\sevensize $^{9}$ Perinotto \& Corradi (1998)}

\end{minipage}
\end{table*}

\vspace{15cm}
\begin{table*}
\begin{minipage}{120mm}
\label{tbl1}
\caption{Nebular parameters derived for $wels$-PNe (see the text for details)} 
\vspace{1cm}
\begin{tabular}{@{}lrrrrrrr} 
\hline
\hline
PNG &  $T_{d}$ & $M_{d}$ & $m_{d}/m_{g}$ & $ L_{IR}$ & IRE & Exc. \\
    &        (K)     & (10$^{-4}$M$_{\sun}$)& (10$^{-3}$) & (L$_{\sun}$)&   & class \\
\hline
000.4+04.4$^{1}$  & 82  & 1.84$\pm$0.89 & --  &  539$\pm$258&  --  &-  & \\
000.7-02.7   &    --    &  --   & -- &   --   & --   & 8 & \\ 
000.7+04.7   &    95  & 3.32$\pm$1.8 &8.89$\pm$1.5  & 2032$\pm$1098& 5.10$\pm$0.72 &2  & \\
000.9-02.0   &    --  &  --   & --  &--   & --   &3  & \\
001.7-04.6   &    93  & 1.34$\pm$0.57 &4.8$\pm$0.83  & 735$\pm$304& 3.58$\pm$0.5 &5  & \\
002.0-06.2   &   90  & 1.60$\pm$0.67 &--  & 749$\pm$305  & -- & - & \\
002.0-13.4   &    116 & 0.55$\pm$0.23 &4.7$\pm$0.82  & 921$\pm$373 & 3.18$\pm$0.45 & 1 & \\ 
002.6+08.1   &    81  & 2.26$\pm$1.0 &18.8$\pm$3.2  & 623$\pm$270 & 4.86$\pm$0.69 & 3 & \\
003.2-04.4   &    --  &  --   &  -- & --   & --   &3  & \\
003.7-04.6   &    84  & 1.62$\pm$0.68 &1.17$\pm$0.2 &  537$\pm$323 & 3.72$\pm$0.55 &9  & \\
             &                 &     &       &        &     &    \\
003.9-14.9   &  127 & 0.37$\pm$0.15 &1.04$\pm$0.19  & 972$\pm$390 & 2.55$\pm$0.36 & - & \\
004.2-04.3   &    --  & -- & -- & --  & --       & 3 & \\
004.6+06.0   &   77  & 2.45$\pm$1.1 &20$\pm$3.6  & 525$\pm$229 & 4.63$\pm$0.66 & 1 & \\
006.4+02.0   &   100 & 2.38$\pm$ 1.0 &4.4$\pm$0.74 & 1882$\pm$775& 3.95$\pm$0.56 & 2 & \\
007.0-06.8   &   94  & 2.22$\pm$0.93 &3.2$\pm$0.57  & 1287$\pm$520& 3.44$\pm$0.49 &2  & \\
007.8-04.4   &   85  & 2.85$\pm$1.2 &32$\pm$5.5  & 1002$\pm$412& 5.73$\pm$0.81 & - & \\
008.1-04.7   &   81  & 2.51$\pm$1.1 & -- & 3451$\pm$1590& 5.70$\pm$0.71  & 2 & \\
009.6-10.6   &   80  & 1.58$\pm$0.61  &--& 410$\pm$241& 4.96$\pm$0.7 & 8  & \\
010.8-01.8   &   78  & 2.78$\pm$1.2 & 11.9$\pm$2.28 &  636$\pm$263&2.55$\pm$0.36  & 3 & \\ 
011.7-00.6   &   70  & 5.16$\pm$2.3 & --  &  686$\pm$309& --   & -  & \\
             &                 &     &       &        &     &   & \\
012.5-09.8   & --  & --    & --&  --    & --   &6  & \\
013.7-10.6  & 81  &  -- & --  &   --   & -- & 6  & \\
014.2+03.8  & 97  & --  & --  &  --    & --   & - & \\
014.4-06.1$^{1}$ & 75  & 2.28$\pm$1.1  &-- &  428$\pm$220&  --  & -  & \\
014.3-05.5  & --  & --    & -- &  --    & --   & - & \\ 
016.4-01.9   & 91  & 1.34$\pm$0.76 &11$\pm$2.0   & 662$\pm$369 & 5.03$\pm$0.71 & 11& \\
019.4-05.3   & 114 & 3.2$\pm$ 1.4 & 6.1$\pm$1.0 & 4872$\pm$2107& 5.15$\pm$0.73 & 3 & \\ 
025.8-17.9   & 94  & 0.49$\pm$0.2 &0.8$\pm$0.13  & 284$\pm$113 & 1.73$\pm$0.25 & 10& \\
034.6+11.8   & 120 & 0.77$\pm$0.33 &1.2$\pm$0.22 & 1530$\pm$622& 2.57$\pm$0.36 & 2 & \\
038.2+12.0   & 110 & 1.05$\pm$0.52 &2.0$\pm$0.35 & 1342$\pm$654& 3.51$\pm$0.49 & 1 & \\
             &                 &     &       &        &     &   & \\
046.4-04.1   & 94  & 1.81$\pm$0.78 &1.01$\pm$0.17 & 1052$\pm$446& 2.3$\pm$0.34  & 5 & \\
051.9-03.8   & 92  & 1.49$\pm$0.65 &9.3$\pm$1.5 & 777$\pm$335& 4.54$\pm$0.64 & 6 & \\
054.1-12.1   & 89  & 1.01$\pm$0.42 & 4.3$\pm$0.7 &  445$\pm$175& 1.78$\pm$0.25 & 2 & \\
055.5-00.5   & 114 & 0.75$\pm$0.35 & 2.44$\pm$0.44 & 1143$\pm$511& 4.93$\pm$0.70 & 2 & \\
057.2-08.9   & 84  & 1.29$\pm$0.55 & 5.4$\pm$0.90& 427$\pm$118& 1.83$\pm$0.26 & 5 & \\
058.3-10.9   & 144 & 0.6$\pm$0.28 &1.0$\pm$0.2  &2953$\pm$1343& 2.09$\pm$0.3 & 6 & \\
068.7+14.8   &  -- & --    & --&  --   & --    & 2 & \\
081.2-14.9$^{3}$&  -- & -- &-- &   --   & --   & 12 & \\
096.4+29.9   &  93  & 1.28$\pm$0.57 & 4.8$\pm$0.8 & 707$\pm$304& --   & 7 & \\
159.0-15.1   &  90  & 0.56$\pm$0.30 &0.51$\pm$0.11 & 266$\pm$152& 1.61$\pm$0.21 & 10 & \\
             &                 &     &       &        &     &   & \\
190.3-17.7   &  85  & 0.63$\pm$0.25 &1.0$\pm$0.17 & 222$\pm$87 & 1.36$\pm$0.19 & 4  & \\
194.2+02.5   &  106 & 0.65$\pm$0.28 &1.3$\pm$0.2 & 690$\pm$286 & 2.10$\pm$0.29 & 9 & \\
208.5+33.2  &   79  & 1.03$\pm$0.40 & $\ge$ 30$\pm$14 &2503$\pm$1001& --   & - \\
221.0-01.4   &   -- & --   & -- &  --   & --   & -  & \\
221.3-12.3   &  106 & 0.4$\pm$0.17 &0.69$\pm$0.12 & 421$\pm$170& 1.17$\pm$0.17 & 10& \\
253.9+05.7   &  90  & 1.74$\pm$0.73 &6.2$\pm$1.1 & 812$\pm$332& 3.7$\pm$0.52 & 4 & \\
258.1-00.3   &  115 & 0.41$\pm$0.18 &0.79$\pm$0.1 & 663$\pm$279& 1.83$\pm$0.26 & 3 & \\ 
264.4-12.7   &  109 & 0.35$\pm$0.15 &1.1$\pm$0.2 & 431$\pm$176& 1.34$\pm$0.19 & 2 & \\
274.6+02.1   &  94 & 1.02$\pm$0.45  &4.4$\pm$0.7&590$\pm$259& 2.39$\pm$0.34 & 3 & \\
289.8+07.7   &  --  & -- &  --& -- & --  & 10 & \\
             &                 &     &       &        &     &   & \\

\hline
\end{tabular}
\end{minipage}
\end{table*}

\begin{table*}
\begin{minipage}{120mm}
\label{tbl3}
\begin{tabular}{@{}lrrrrrrr}
\hline
\hline
PNG & $T_{d}$ & $M_{d}$ & $m_{d}/m_{g}$ & $ L_{IR}$ & IRE & Exc. \\
    &         (K)     & (10$^{-4}$M$_{\sun}$)& (10$^{-3}$) & (L$_{\sun}$)&   & class \\
\hline 

315.1-13.0 &117 & 1.27$\pm$0.535  & 11.9$\pm$2.0& 2212$\pm$908& 5.26$\pm$0.74 &10 & \\
316.1+08.4 & 84 & 2.16$\pm$0.88  & 22.7$\pm$3.5& 714$\pm$289 & 7.1$\pm$1.0 &10 & \\
329.9+03.7 & 78 & --    & -- &  --   & --   & -  & \\  
331.3+16.8 &108 & 0.23$\pm$0.09&1.2$\pm$0.2 &267$\pm$107 & 1.3$\pm$0.19 &10 & \\
331.8-02.3 & --  &  --   &-- &  --   & --   & - \\
333.4-04.3 & 79 &0.83$\pm$0.37  & -- & 203$\pm$110& --   &-  & \\
341.8+05.4  & 78 & 2.78$\pm$1.2 &11.9$\pm$2.3 & 633$\pm$263& 2.29$\pm$0.6 & 8 & \\
343.9-05.8$^{1}$ &  -- & -- & --  &   --   & --   & 10 & \\
351.7-06.6$^{1}$ &  -- & -- & --  &   --   & --   & 3  & \\
             &                 &     &       &        &     &   & \\
355.9-04.2 & 86 & 3.15$\pm$1.5 &10.5$\pm$2.3 & 1172$\pm$516& 4.1$\pm$0.57 & 1 & \\
356.2-04.4 & 105& 1.32$\pm$0.56 &3.42$\pm$0.6 & 1338$\pm$548& 2.15$\pm$0.3 & 7 & \\
356.7-04.8 & 82 & 0.8$\pm$0.33 &7.1$\pm$1.1& 234$\pm$96& 1.94$\pm$0.27 & 9 & \\
356.9+04.4 & 89 & 9.90$\pm$4.5 &35.12$\pm$4.71& 4375$\pm$1899& 8.58$\pm$1.2 & 7 & \\
357.1+03.6 &  --&  --  &-- &   --   & --   & 2 & \\
357.2-04.5 &100 & 0.68$\pm$0.3 &0.62$\pm$0.11 &  529$\pm$285& 1.72$\pm$0.21 & 3  & \\
358.0-04.6 & -- &  --  &-- &   --   & --   & -  & \\ 
358.9+03.3 &97  & 3.71$\pm$1.6 &1.3$\pm$0.2 & 2523$\pm$1065& 4.71$\pm$0.67 & -  & \\
\hline 
\end{tabular}
\\
\end{minipage}
\end{table*}

\vspace{15cm}
\begin{table*}
\begin{minipage}{120mm}
\label{tbl4}
\caption{Nebular parameters derived for normal PNe (see the text for details)} 
\vspace{1cm}
\begin{tabular}{@{}lrrrrrrr}
\hline
\hline
PNG &  $T_{d}$ & $M_{d}$ & $m_{d}/m_{g}$ & $ L_{IR}$ & IRE & Exc. \\
    &       (K)     & (10$^{-4}$M$_{\sun}$)& (10$^{-3}$) & (L$_{\sun}$)&   & class \\
\hline
000.1+17.2 &  98 &  2.25$\pm$1.15 & 5.0$\pm$0.8 & 1609$\pm$816 & 4.29$\pm$0.61 & 1 &  \\ 
002.2-02.7 & 240 &  0.58$\pm$0.29 &0.42$\pm$0.09& 3691$\pm$1692 & 11.6$\pm$1.64& 3 &  \\
002.4-03.7 & 105 &  1.38$\pm$0.62 &11.1$\pm$2.1 & 1392$\pm$612 & 5.23$\pm$0.74 & - &  \\
002.7-52.4  & --  &  --  & -- &  --      & --   & 2  &  \\
008.2+06.8$^{4}$ & 147 & 1.06$\pm$0.56 &2.7$\pm$0.47 & 5771$\pm$3001  &13.7$\pm$1.9 & 1 &  \\
010.8+18.0 &  95 &   --  & -- &   --     & -- & - &  \\
010.7-06.4 & 143 &  0.31$\pm$0.13 & 0.40$\pm$0.72 & 1459$\pm$580 & 1.71$\pm$0.24 & 5 &  \\
011.7-06.6 & --  &   --  &  -- & -- & --  & - &  \\   
014.6-04.3 & 87  & 2.35$\pm$0.90&3.7$\pm$0.68 & 925$\pm$376 & 2.45$\pm$0.35& 7 &  \\ 
015.9+03.3 & 100 &   --  &  --  & --  & -- & 1 &  \\ 
            &                 &     &       &        &        & \\
017.3-21.9 & --  &  --   &  --      & -- & -- & 1 &  \\
017.6-10.2 & 81  & 0.29$\pm$0.12 & 25.3$\pm$6.7 & 80$\pm$32 & 7.83$\pm$1.11 & 12&  \\
025.3+40.8 & 98  & 0.88$\pm$0.32 & -- & 632$\pm$217   & --   & 6 &  \\
025.4-04.7 & 75  & 0.70$\pm$0.2 &4.6$\pm$0.77 & 13.22$\pm$5.2 & 0.82$\pm$0.11 &10 &  \\
027.6-09.6 & 106 & 0.81$\pm$0.33 &1.1$\pm$0.2 & 863$\pm$344 & 1.62$\pm$0.22 & 4 &  \\
034.5-06.7 & 59  & 3.74$\pm$1.5 &4.1$\pm$0.7 & 212$\pm$85 & 1.62$\pm$0.23 & 9 &  \\
035.9-01.1 & 81  & 0.10$\pm$0.06 & -- & 24$\pm$13 & 0.93$\pm$0.13& 3 &  \\
036.1-57.1  & --  &   --  & -- &  --     & --   & 2 &  \\
037.7-34.5$^{5}$ & 86 & 1.32$\pm$0.55 &5.2$\pm$1 & 491$\pm$198  & 1.92$\pm$0.27 & 7 &  \\
042.9-06.9 & 166 & 0.25$\pm$0.10 &1.6$\pm$0.27 & 2474$\pm$996 & 4.33$\pm$0.61 & 3 &  \\
            &                 &     &       &        &     &   & \\
043.1+37.7 &  90  & 1.17$\pm$0.48 &2.29$\pm$0.38 & 548$\pm$219 & 2.17$\pm$0.3 & 4 &  \\
045.4-02.7 &  132 & 2.23$\pm$1.02 &15.5$\pm$2.7 & 7087$\pm$3152  & 45.9$\pm$6.5& 3 &  \\
045.7-04.5 &  88  & 0.37$\pm$0.16 & 7.9$\pm$1.4& 153$\pm$63 & 3.1$\pm$0.44 & 12 &  \\
047.0+42.4 &  --  &   --  & -- &  --      & --   & - &  \\  
049.4+02.4 &  62  &0.75$\pm$0.34 & -- &  54$\pm$24   & 29.9$\pm$4.2& 1 &  \\
051.0+02.8 &  132 & --& -- & -- & --  & 3 &  \\ 
053.8-03.0 &  --  &  --   &-- & --       & --  & 1 &  \\
055.4+16.0 &  --  &  --   &-- & --       & --  & 2 &  \\
060.1-07.7 &  94  & 1.56$\pm$0.64 &5.75$\pm$0.9 & 903$\pm$365 & 2.1$\pm$0.29&11 &  \\
060.8-03.6 &  63  & 0.05$\pm$0.02 & -- &  4$\pm$2 & -- & 11&  \\
           &    &     &       &        &     &   &  \\
061.0+08.0 & 95  & 0.24$\pm$0.10 &7.1$\pm$1.1 &147$\pm$61 & 4.1$\pm$0.57&12+&  \\
063.1+13.9 & 65  & 1.03$\pm$0.42 &1.3$\pm$0.2 &95$\pm$38 & 1.45$\pm$0.21 & 10 & \\ 
064.6+48.2 & 76  & 0.48$\pm$0.19 &17.8$\pm$2.8 & 96$\pm$38 & 3.30$\pm$0.47 & 11&  \\
072.7-17.1 & --  &   --  &-- &  --      & --  & - &  \\
077.6+14.7 & --  &  --   &-- &  --      & --  & 3 &  \\ 
082.5+11.3 & --  &   --  &-- &   --     & --   & 1 &  \\
083.5+12.7 & 93  & 0.67$\pm$0.07 & 1.97$\pm$0.3& 368$\pm$14.9 & 2.1$\pm$0.29 & 8&  \\
089.3-02.2 & 101 & 1.41$\pm$0.61  & -- & 1175$\pm$611  & 11.7$\pm$2.1 &- &  \\
093.4+05.4 & 83  & 0.47$\pm$0.2 & 17.8$\pm$0.39& 146$\pm$59 & 4.26$\pm$0.6 & 11 &  \\
095.2+00.7 & 127 & 1.08$\pm$0.52 & 1.0$\pm$0.2 & 2827$\pm$1361 & 2.45$\pm$0.35 & 3&  \\
            &                 &     &       &        &     &   & \\
102.9-02.3$^{6}$ &  -- &  --&-- &  --      & --   &3&  \\
107.8+02.3  & 89 & 0.64$\pm$0.27 &5.1$\pm$0.83& 284$\pm$118 & 2.13$\pm$0.3 & 9 &  \\
114.0-04.6 &  -- &  --   & -- &  --      & --   & - &  \\
123.6+34.5 & 93  & 0.46$\pm$0.05 &4.3$\pm$0.71& 254$\pm$15 & 2.23$\pm$0.32 & 3 &  \\
147.4-02.3 & 95  & 0.68$\pm$0.29 &2.9$\pm$0.46& 414$\pm$175 & 0.79$\pm$0.11 & 3 &  \\
148.4+57.0$^{7}$& --  &  --   &--& --     & --   & 8 &  \\
165.5-06.5  & 93  &  --   &  --&  --     & -- & 3 &  \\
165.5-15.2  & 58  & 0.17$\pm$0.07& -- &  9$\pm$4    & --   & - &  \\
166.1+10.4  & 104 & 0.60$\pm$0.24 &1.2$\pm$0.2 & 574$\pm$230 & 1.73$\pm$0.25 & - &  \\
167.4-09.1  & 115 & 0.26$\pm$0.11 & 1.2$\pm$0.2  & 408$\pm$170 & 1.74$\pm$0.25 & 1 &  \\
\hline
\end{tabular}
\end{minipage}
\end{table*}

\begin{table*}
\begin{minipage}{120mm}
\label{tbl4}
\begin{tabular}{@{}lrrrrrrr}
\hline
\hline
PNG &  $T_{d}$ & $M_{d}$ & $m_{d}/m_{g}$ & $ L_{IR}$ & IRE & Exc. \\
    &         (K)     & (10$^{-4}$M$_{\sun}$)& (10$^{-3}$) & (L$_{\sun}$)&   & class \\
\hline 
170.3+15.8 & 100 &  0.11$\pm$0.05 & -- &  84$\pm$35  & --   &12+& \\
193.6-09.5 &  -- &  --   &--&  --      & --   & 8 &  \\
196.6-10.9 & 101 &0.38$\pm$0.16 & 0.8$\pm$0.13& 316$\pm$127& 3.52$\pm$0.5 &12 &  \\
197.8-03.3 &  -- &  --   & -- & --      & --   &-  &  \\
197.8+17.3 & 78 &  0.48$\pm$0.02& 1.1$\pm$0.2 & 110$\pm$45 & 1.27$\pm$0.18 & 8 &  \\
204.1+04.7 & --  &   --  &   -- &--     & --   & 1 &  \\ 
206.4-40.5 & 83  &  1.6$\pm$0.5& -- & 500$\pm$265 & 1.45$\pm$0.20 & 8 &  \\ 
214.9+07.8 & 94  &  0.04$\pm$0.02 &2.0$\pm$0.4 & 25$\pm$10 & 1.13$\pm$0.16 &12 &  \\
215.2-24.2 & 124 &  0.67$\pm$0.28 &1.8$\pm$0.3 & 1553$\pm$637 & 2.49$\pm$0.35 & 1 & \\
215.6+03.6 & 59  &  0.67$\pm$0.34 &2.8$\pm$0.5 & 38$\pm$18 & 1.57$\pm$0.22 & 9 &  \\
            &                 &     &       &        &     &   & \\
219.1+31.2$^{8}$ & --  &  --&  --  &  --      & --  & 1  &  \\
220.3-53.9 & 71  & -- & -- & --    & --   &12 &  \\
231.8+04.1 & 63  & 0.42$\pm$0.19 & 0.7$\pm$0.2 & 33$\pm$13 & 1.1$\pm$0.15 & 5 &  \\
233.5-16.3 & 90  & 0.13$\pm$0.06 &7.6$\pm$1.2 & 61$\pm$28 & 3.36$\pm$0.47&12+ &  \\
234.9-01.4 & 109 & 0.37$\pm$0.17 & 0.75$\pm$0.2 & 456$\pm$271 & 1.57$\pm$0.24 & 1 &  \\
238.0+34.8 & --  &   --  & -- &  --     & --  &1 &   \\  
239.6+13.9 & 98  & 0.10$\pm$0.04 &3.2$\pm$0.6 & 67$\pm$27 & 2.17$\pm$0.3 &12 &  \\
241.0+02.3 & --  &   --  & -- & --      & --   & 11&  \\
242.6-11.6 & 69  & 0.63$\pm$0.26 &1.4$\pm$0.2 & 79$\pm$32 & 0.74$\pm$0.1 & 3 &  \\
245.4+01.6 & 64  & 2.95$\pm$1.3 &4.1$\pm$0.7 & 251$\pm$105 & 3.01$\pm$0.4 & 9 &  \\
            &                 &     &       &        &     &   & \\
248.7+29.5 & --  &   -- &--  &   --     & --  & - &  \\
253.5+10.7 & 105 & 0.03$\pm$0.01&--& 25$\pm$13 & --   &12 &  \\
255.3-59.6 & --  &   -- & -- &   --     & --  & - &  \\
261.0+32.0 & 87  & 0.50$\pm$0.27& -- & 197$\pm$105& --   & 9 &  \\
261.9+08.5 & 78  & 0.13$\pm$0.04 &5.0$\pm$0.9 & 30$\pm$12 & 1.07$\pm$0.15 & 10&  \\  
263.2+00.4 &  -- &  --   &-- &  --      & --   & 1 &  \\
272.1+12.3 & 59  & 1.67$\pm$0.5  & -- & 94$\pm$43    & --   & 5 &  \\
277.1-03.8 & 70  &  0.19$\pm$0.07&0.38$\pm$0.06 & 25$\pm$10 & 1.15$\pm$0.16 & 11&  \\
279.6-03.1$^{9}$ & 90 & 0.38$\pm$0.15 &3.4$\pm$0.5 & 176$\pm$71& 1.48$\pm$0.2& 10&  \\
283.6+25.3 & --  &   --  &  --      & -- & --  & 1 &  \\
            &                 &     &       &        &     &   & \\
285.7-14.9 & 89  &0.32$\pm$0.13&3.6$\pm$0.96 & 144$\pm$68 & 1.50$\pm$0.2 & 10&  \\
294.1+14.4 & --  &   --  & -- &  --      & --  & 3 &  \\
294.1+43.6 & 98  & 0.07$\pm$0.03 &-- & 49$\pm$20 & --   & 12+&  \\
327.8+10.0 & 88  & 1.65$\pm$0.7 &16.8$\pm$2.8 & 691$\pm$276 & 2.52$\pm$0.36 & 6 &  \\
341.6+13.7 &  76 & 0.23$\pm$0.1 &1.1$\pm$0.2 & 47$\pm$19 & 0.20$\pm$0.03 & 11&  \\
345.4+00.1 & --  &   --  &-- &  --      & --   &3  &  \\
303.6+40.0 & --  &   --  &-- &  --      & --  & 1 &  \\
305.1+01.4 & 139 &   --  &-- &  --      & --  & 1 &  \\
308.6-12.2 & --  &   --  &-- &  --      & --  & 1 &  \\
339.9+88.4 & --  &   --  &-- &  --      & --  & - &  \\
           &                 &     &       &     &     &   &  \\
310.3+24.7 & --  &  --   &-- &  --      & --   & 12 &  \\
318.4+41.4 & --  &  --   &-- &  --      & --   & - &  \\
320.1-09.6 & 102 & 3.71$\pm$1.4 & 2.8$\pm$0.8 & 3245$\pm$1880 & 9.30$\pm$1.5 & - &  \\
320.3-28.8 & 102 & 0.45$\pm$0.13 &1.5$\pm$0.5 & 391$\pm$203 & 2.00$\pm$0.3 & 3 &  \\
325.8-12.8 & 171 & 0.15$\pm$0.07 &1.1$\pm$0.2  & 1963$\pm$823 & 9.28$\pm$1.1 & 1 &  \\
331.4-03.5 &  -- &  -- & --  &  --      & --  & 1 &   \\
336.3-05.6 & 100 & 0.40$\pm$0.12 & -- & 319$\pm$153 & 2.06$\pm$0.32 & 9 &  \\
345.5+15.1 &  -- &  --   & -- &  --      & --   & - &  \\
349.5+01.0 & 76  & 4.67$\pm$0.2 &17.8$\pm$2.8  & 937$\pm$379 & 4.04$\pm$0.65 & 10 & \\  
357.6+01.7 & --  & --  & -- & --  & -- & 3 &  \\
\hline  
\end{tabular} 
\\
\end{minipage}
\end{table*}

\vspace{2.0cm} \protect \begin{figure*} 
\includegraphics[width=\textwidth]{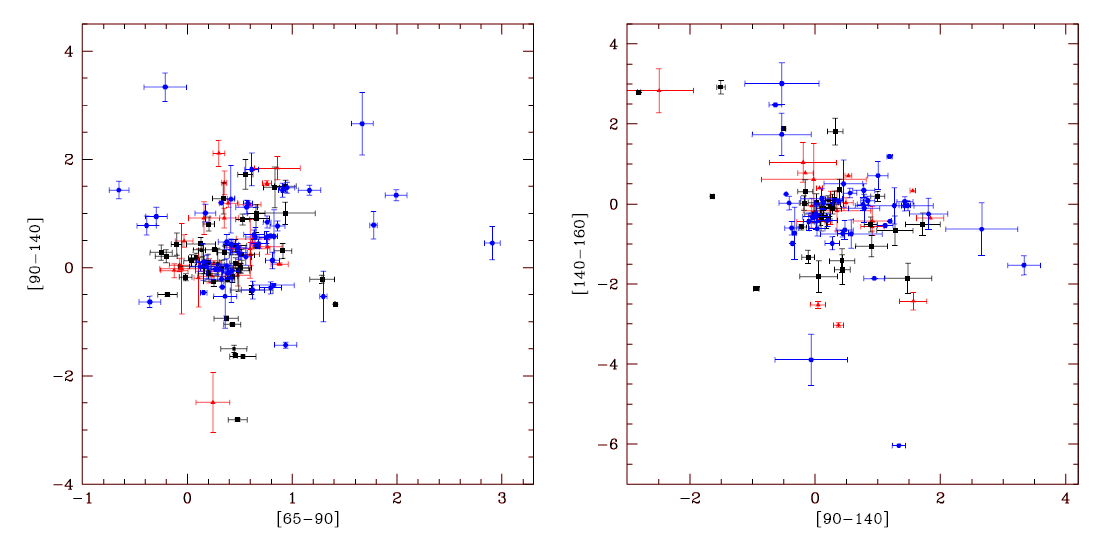}
\caption{Far-IR colour-colour diagrams of PNe using $Akari$ measurements. Data description as given in Fig. 1. See the text for details.} 
\end{figure*}

\vspace{2.0cm} \protect \begin{figure*} 
\includegraphics[width=\textwidth]{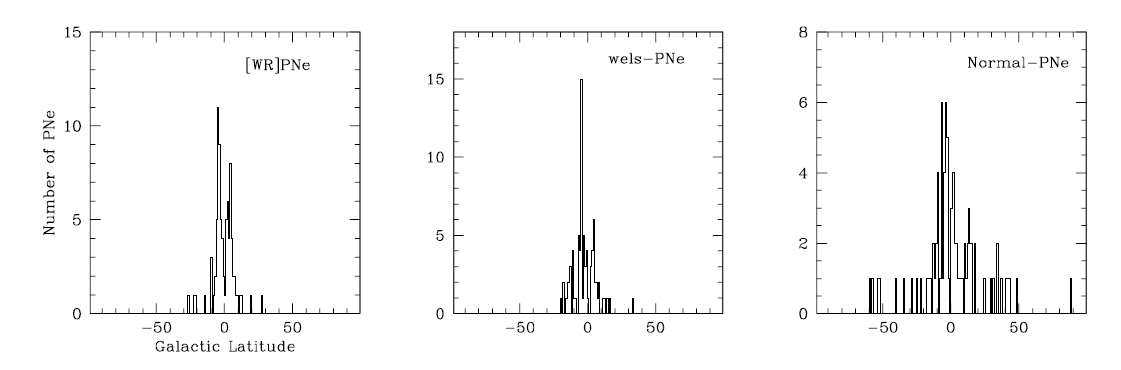}
\vspace{1.0cm}
\caption{Histograms showing Galactic distribution of [WR] PNe, $wels$-PNe and normal PNe.  See the text for details.} 
\end{figure*}


 \section{Gas and Dust properties}
  The properties of gas and dust components of [WR] PNe, $wels$-PNe and normal-PNe were derived using IRAS photometric
  data and are given in Table 2, Table 3 and in Table 4 respectively. In this section we discuss these properties 
  in more details.

 \noindent
 \subsection{Dust Colour Temperature}
 
 The colour temperature of thermal equilibrium dust in a PN can be derived using its continuum fluxes at
 $IRAS$ 25- and 60$\mu$m bands. Though there is a distribution of dust temperature in a PN due different
 radial distances of grains from the heating central star and also due to the grain size distribution,
 the resulting IR spectrum from a PN reasonably indicates a characteristic dust temperature of the PN
 (Stasinska \& Szczerba 1999). We derived the characteristic dust colour temperatures, 
 or simply the colour temperatures ($T_{d}$) of PNe using their  $IRAS$ 25- and 60$\mu$m fluxes. The
 ratio of fluxes emitted from dust at wavelengths $\lambda_{1}$ and $\lambda_{2}$ in the optically thin 
 limit is:   
 
 \begin{equation}
 F_{\lambda_{1}}/ F_{\lambda_{2}} =  (\lambda_{2}/\lambda_{1})^{\alpha} \times B_{\lambda_{1}}(T_{d})/ B_ {\lambda_{2}}(T_{d}) 
 \end{equation}   

\noindent 
 where, $B_{\lambda_{1}}(T_{d})$ and $B_{\lambda_{2}}(T_{d})$ are the values of Planckian function at $\lambda_{1}$ and 
 $\lambda_{2}$ for $T_{d}$. We fitted the $IRAS$ fluxes using this function with an emissivity index 
 $\alpha$ = 1 to obtain $T_{d}$. The dust colour temperatures were computed for PNe if their $IRAS$ 25- and 60$\mu$m 
 fluxes are available with error estimates. The assumed value of $\alpha$ is typical for micron-sized grains. The colour
 temperatures for PNe were earlier derived by Zhang \& Kwok (1993) using $IRAS$ fluxes with a blackbody model 
 ($\alpha$ = 0) fit. However, dust do not emit like blackbodies and as discussed earlier, a modified black body curve with  $\alpha$ = 1 represents the
 continuum emission from circumstellar dust better. Increasing the value of $\alpha$ will result a lower colour temperature for dust. The error in dust colour temperature is due to the errors
 in the $IRAS$ fluxes for a given $\alpha$. This error was examined by fitting the curve to extreme ranges in the flux 
 ratios. The error in T$_{d}$ estimation never exceeded 10$\%$ and we have kept this as the error in $T_{d}$ estimation. 
 The mean dust colour temperatures of 
 [WR] PNe, $wels$-PNe and of normal-PNe were derived and are found to be quite similar taking their respective dispersions (or the observed ranges) into account (see Table 5).
 
 A plot between the dust colour temperatures of PNe and their ages will be an useful tool to see if there is a change
 in dust temperature as a PN evolves. A robust, distance-independent parameter to indicate the age of a PN is the 
 nebular H${\beta}$ surface brightness ($S_{H\beta}$; Stasinska \& Szczerba 1999; Gorny \& Tylenda 2000).  $S_{H\beta}$ 
 decreases as a PN ages. If the interstellar reddening corrected nebular H$\beta$ flux is $F_{H\beta}$ and the
 angular radius of the optical nebula is $\theta$ then $S_{H\beta}$ is defined as: \\

 \begin{equation}
 S_{H\beta} = F_{H\beta}/\pi \theta^{2} 
 \end{equation}

 PNe are often non-spherical in shape. The values of $\theta$ for all PNe in our sample are calculated from their 
 angular sizes of major and minor axes given in FPB16, by equating the area. We could not
 estimate the error in $S_{H\beta}$ as the errors in angular extends of optical nebulae are not available; however, it 
 is unlikely that the error in $S_{H\beta}$ will be significant enough to change the trends seen in our analysis. Fig 7a 
 shows the plot of $T_{d}$ against $S_{H\beta}$ for [WR] PNe, $wels$-PNe and in normal PNe, indicating that the dust 
 colour temperature decreases steadily with age for PNe. This is similar to the earlier finding by Gorny et al. (2001) for 
 [WR] and non-[WR] PNe and also by Pottasch et al. (1984). However, we resolve [WR] PNe, $wels$-PNe and normal-PNe in our 
 dust colour temperature plot against age. To quantitatively find the strength of correlation between $T_{d}$ and $S_{H\beta}$,
 we estimate the Pearson correlation coefficient with an associated probability ($r,p$) for the three groups of PNe
 and they are also shown in the figure. The values of ($r,p$) indicates that the correlation of $T_d$ with age is
 quite tight for all the three groups of PNe. Fig 6a can be understood in terms of expansion of the PN which 
 dilutes the stellar radiation field inside the nebula and as a result of this the grains are less heated. The figure 
 also shows that [WR] PNe and $wels$-PNe are not observed at large ages like normal PNe. [WR] PNe are not observed 
 below $S_{H\beta}$ = $3\times10^{-4}$ and $wels$-PNe are seen only above $S_{H\beta}$ = $10^{-3}$, however about 20$\%$ 
 of normal PNe are seen with lower $S_{H\beta}$ values, down to $2\times10^{-5}$. This may show that [WR] PNe and $wels$-PNe
 are relatively younger than a good fraction of normal PNe. The highest values of $S_{H\beta}$ for $wels$-PNe and normal 
 PNe are same while one [WR]PN (Hen 2-113; [WC10]) shows a value which is $\sim$ 2.5 times more than this. The minimum 
 dust temperature observed for [WR] PNe is 63K and for normal PNe it is 59K whereas $wels$-PNe show a significantly 
 higher value of 77. 
 
 Fig 6b shows a plot of $T_{d}$ $vs$ $S_{H\beta}$ for different subtypes of [WR] PNe; viz. [WCE] ([WO1] to
 [WC4]), [WCL] ([WC9] to [WC11]) (See AN03) and the intermediate ones, [WCg], ([WC5] to [WC8]). The figure brings out 
 a trend of decrease in $T_{d}$ from [WCE] to [WCg] to [WCL]. Together, there is also a continuity in the plot, whereas
 in Fig 6a each group of PNe shows such a continuity.

\vspace{2.0cm} \protect \begin{figure*}
\includegraphics[width=\textwidth]{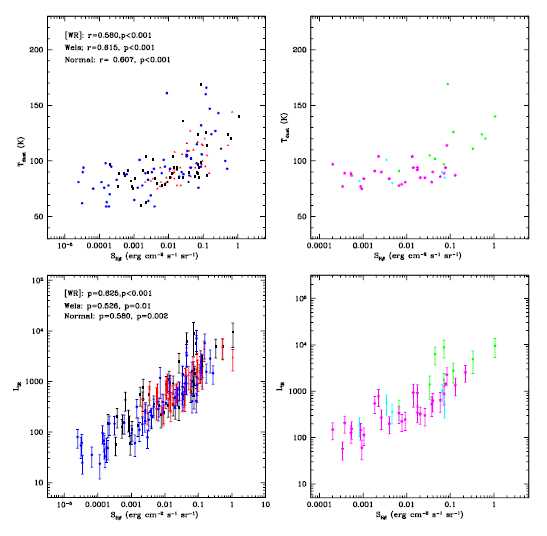}
\caption{$Top$: a) Dust colour temperature is plotted against $S_{H\beta}$ for [WR] PNe, $wels$-PNe and normal PNe (left) 
and b) for the subtypes of [WR] PNe (right). $Bottom$: 
c) IR luminosity for [WR] PNe, $wels$-PNe and normal PNe (left); d) IR luminosity for the subtypes of [WR] PNe (right).
 Data description as shown in Fig 1. See the text for details.} 
\end{figure*}
  
 \noindent
 \subsection{IR luminosity and IR excess} 
 
 We estimate the IR luminosity ($L_{IR}$) of a PN as the total IR continuum energy emitted by thermal
 equilibrium dust. Hence $L_{IR}$ does not include the emission due to dust features and the emission due to the
 hot dust component. $L_{IR}$ of a PN is calculated using its $IRAS$ fluxes at 25- 
 and 60$\mu$m bands. We fitted a modified blackbody flux distribution function with an emissivity index $\alpha$ = 1, 
 and integrated the whole curve to get $L_{IR}$. Error in $L_{IR}$ was estimated from the error in the integrated flux and 
 from (and dominated by) the error in distance measurement which is given in FPB16. The 
 IR excess (IRE) represents the excess amount of $L_{IR}$ in terms of what can be accounted by absorption of Ly$\alpha$ 
 radiation produced in the nebula. The IRE of a PN was calculated from its $L_{IR}$ and its reddening corrected nebular 
 H$\beta$ flux using the relation given in Stasinska \& Szczerba (1999; equation 6). The error in IRE was estimated 
 from the errors in respective parameters used to calculate IRE.  A plot on the variation of $L_{IR}$ and IRE of PNe 
 against their $S_{H\beta}$ will be useful to address how these parameters change as the PN evolves. Fig 6c shows the 
 variation of  $L_{IR}$ with $S_{H\beta}$ for [WR] PNe, $wels$-PNe and normal-PNe along with their respective ($r,p$) 
 values showing the correlation strengths. The figure displays a tight correlation between $L_{IR}$ with $S_{H\beta}$ 
 for all the three groups of PNe, implying that $L_{IR}$ decreases strongly with nebular age for all PNe. The variation of $L_{IR}$ 
 with $S_{H\beta}$ for the subgroups of [WR] PNe can be seen in Fig 6d which shows a decrease in $L_{IR}$ for PNe with 
 late type to early type [WR] central stars. However, as seen in Fig 7a, IRE is not correlated with $S_{H\beta}$ for all 
 the PNe, which was also found earlier by Stasinska \& Szczerba (1999) for their sample. This is in contrast with the 
 suggestion given by Pottasch et al. (1984) who had studied $L_{IR}$ and IRE of PNe and proposed that  younger PNe have higher IRE. As a PN expands to a larger size and the central star is becoming hotter,
 its $L_{IR}$ drops steadily keeping the IRE constant.  
   
 The average values and dispersions of $L_{IR}$ and IRE for the three groups of PNe were computed and are compared in
 Table 5. As it can be seen in the table, the average value of $L_{IR}$ for [WR] PNe, $wels$-PNe and normal PNe are very similar. However, the mean values of 
 IRE for the three groups of PNe show that $wels$-PNe and normal PNe have quite similar values, while [WR] PNe show a tendency towards a larger mean value. Among [WR] PNe He 2-113, M 2-43, He 2-459, IRAS 21282+5050, He 1-43, SwST 1, 
 K 2-16, BD+30 3639 and IC 5117 show very large values for $L_{IR}$ (IRE) than the other candidates in the group. Their 
 respective values are 9700$L_{\sun}$ ($\ge$ 25), 9498$L_{\sun}$ (7.5), 8834$L_{\sun}$ (14.5), 6689$L_{\sun}$ (--), 6226$L_{\sun}$ (17), 5824 $L_{\sun}$ (15), 5618$L_{\sun}$(--), 4884$L_{\sun}$ (6.75),  4816$L_{\sun}$ (5.2) and 4165$L_{\sun}$(--). 
 All of them belong to the [WCL] subtype. M 1-61 and M 3-38 are the two $wels$-PNe which show large $L_{IR}$ (IRE); 
 their values are 4877$L_{\sun}$ (5.1) and 4375$L_{\sun}$ (8.5). Two normal PNe, namely Vy 2-2 and He 2-260 have their respective
 $L_{IR}$ (IRE) of 7087$L_{\sun}$(15) and 5771$L_{\sun}$(13). The lowest $L_{IR}$ in our sample was seen for a few normal PNe.
 From the equation used to compute IRE (equation 6. in Stasinska \& Szczerba 1999), a weak correlation
 between IRE and $S_{H\beta}$ for all PNe implies that $L_{IR}$ decreases with the nebular $H_{\beta}$ 
 flux for a PN as it expands.
       
\noindent
\subsection{Dust mass and dust-to-gas mass ratio}
    
The dust masses and the dust-to-gas mass ratios of PNe are also examined in order to understand
the nature of the three groups of PNe considered in this study. For an optically thin dust cloud, a criterion 
which is readily fulfilled in and above the mid-IR region for the thermal grains of PNe, the total dust mass can be
obtained from the continuum flux observed at one band where the dust emission dominates. If $m_{\rm d}$ is the total 
dust mass of a PN, it is calculated from the dust temperature and flux at an IR band in the optically thin limit 
using the following relation (Stasinska \& Scczerba 1999): 

\begin{equation}
m_{\rm d}  =  F_{\rm \nu}(\lambda)D^{2}/Q_{abs, \nu} B_{\nu}(T_{d})
\end{equation} 
 
\noindent
 where, $D$ is the distance to the nebula, $F_{\nu}(\lambda)$ is the flux at an IR band (we have taken the flux at 
 $IRAS$ 60$\mu$m which is least contaminated by atomic emission) and $B_{\nu}(T_{d})$ is the Planckian function at 
 frequency $\nu$ (corresponding to 60$\mu$m) for a dust colour temperature of $T_{d}$. $Q_{abs, \nu}(\lambda)$ is the dust
 absorption coefficient at $\lambda$, which has a value of 74 $cm^{2} g^{-1}$ at 60$\mu$m. The error in $m_{\rm d}$ was
 estimated from the errors in respective parameters used to derive $m_{\rm d}$ and is dominated by the error in distance. \\
  
\noindent 
  Dust masses were estimated for all PNe for which $IRAS$ 60$\mu$m fluxes are available with error measurements. Fig 7b shows 
  the plot of dust mass against $S_{H\beta}$ for [WR] PNe, $wels$-PNe and normal-PNe along with their respective ($r,p$) values.
  The figure brings out that the dust mass of a PN is weakly correlated with its age. It shows that the dust is neither destroyed
  nor significantly reduced as a PN evolves, for all the three PNe groups. The mean dust mass values arrived for
  the [WR] PNe, $wels$-PNe and normal PNe are given in Table 5 along with their dispersions. [WR] PNe (and also $wels$-PNe) have
  a tendency towards larger mean dust mass in compared to the normal PNe. It may imply that dust can be created more efficiently 
  by the progenitors of [WR] PNe and $wels$-PNe in compared to the progenitors of normal-PNe. Large IRE for [WR] PNe can also be
  produced when their dust masses are also high. The lowest $m_{d}$ in our sample was seen for a few normal PNe.
    
  The dust-to-gas mass ratios ($m_{d}/m_{g}$) of PNe were calculated using their $IRAS$ fluxes at 25- and 
  60$\mu$m bands. Reddening-corrected nebular H$\beta$ fluxes and the nebular electron temperatures and densities which 
  are required for estimating m$_{d}/m_{g}$ were taken from the literature. We used the relation given by Stasinska
  \& Szczerba (1999; equation 5). The dust temperatures required for this calculation were obtained earlier in Section 4.1. 
  Error in m$_{d}/m_{g}$ was estimated from the errors in the respective parameters used to derive m$_{d}/m_{g}$. The derived 
  values of m$_{d}/m_{g}$ are plotted against $S_{H\beta}$ for the three groups of PNe in Fig 7b and their ($r,p$) values are 
  shown. As it can be seen in the figure, m$_{d}/m_{g}$ is not correlated with the age of the nebula, as also
 was noted by Gorny et al. (2001). These observed trends imply that dust destruction is
  inefficient as a PN evolves. In contrast, Pottasch et al. (1984) suggested from a sample of 46 PNe that 
 m$_{d}/m_{g}$ decreases with nebular radius and hence with nebular age. 
 
  The mean values of m$_{d}/m_{g}$ were estimated for [WR] PNe, $wels$-PNe and normal-PNe which are shown in Table 5 along 
  with their dispersions. All the three groups of PNe show similar mean values, taking the dispersion into account. Gorny et al.
 (2001) also have suggested that there is no significant difference in $m_{d}/m_{g}$ for [WR] PNe and non-[WR] PNe. As the
  mean dust mass tends to have a larger value for [WR] PNe (and $wels$-PNe), similar values of mean $m_{d}/m_{g}$ for all the
  three group of PNe possibly shows that [WR] PNe (and $wels-PNe$) have proportionally larger mean gas mass than normal-PNe. 
  The mean electron densities ($n_{e}$) for [WR] PNe, $wels$-PNe normal-PNe are also similar (Table 5). 

  The chemical nature of the grain (amorphous carbon or amorphous silicate) in PN can also contribute to the estimation 
  of $T_{d}$, $m_{d}$, $m_{d}/m_{g}$, $L_{IR}$ and IRE. To derive the nebular and dust parameters above we have considered 
  a dust emissivity index of $\alpha$ = 1 and $Q_{abs, \nu}(\lambda)$ = 74.38 (Stasinska \& Szczerba 1999), which are more 
  suitable for amorphous carbon grains. If the grain type is chosen to be amorphous silicate then $\alpha$ $\sim$ 2 and 
  $Q_{abs, \nu}(\lambda)$ = 53.45 (Stasinska \& Szczerba 1999); which  will result about 30$\%$ difference in the derived 
  values of $T_{d}$, $m_{d}$ and $m_{d}/m_{g}$ and a few percent decrease in the $L_{IR}$ and IRE. The mean nebular and 
  dust parameters for the three groups of PNe taking the grains to be amorphous silicates are also given in Table 5 along 
  with their respective values derived for amorphous carbon grains. As noted by Stasinska \& Szczerba (1999), $m_{d}/m_{g}$ 
  can be overestimated for PNe that are not fully ionized, however, such cases should represent a small fraction for the 
  low surface brightness PNe. 
 
  Szczerba et al. (2001) noted from their IR spectroscopic study of 16 [WR] PNe that $\sim$ 75$\%$ of them
 show the signatures of PAH emission and hence they can have carbon based dust. We have searched $Spitzer$ 
 IRS spectra for the sample of PNe considered in our study. We found that 18 [WR]PNe, 15 $wels$-PNe and 19 
 normal PNe have their spectra in the $Spitzer$ $Heritage$ $Archive$. We also looked at the $ISO-SWS$ spectra
 where, the 3.3- and 4.6$\mu$ features are either absent or very weak. Our preliminary inspection of $Spitzer$
 IRS spectra shows that 16 [WR]PNe (88\%) and 10 $wels$-PNe (67$\%$) and 10 normal PNe (52$\%$) in our sample
 show clear presence of PAH features. However, [WR]PNe usually show stronger PAH features than the other two
 groups, confirming the trend seen in our photometric study in Section 3.2 that they likely to have brighter
 12$\mu$m $IRAS$ band. These spectra also show that the dust chemistry in most [WR]PNe and $wels$-PNe are 
 carbon-rich. Gorny et al. (2001) argued from their $IRAS$ CCDMs of carbon stars and [WR] PNe that they are
 evolutionarily connected. The distribution of nebular C/O ratio of [WR] PNe is similar with the other PNe 
 as found by Gorny \& Stansinska (1995). However, many of the $wels$ appear in Mendez (1991) are oxygen rich
 stars and are related to variable stars with oxygen-rich circumstellar shells (Acker, Gorny \& Cuisinier
 1996). The grain chemistry is not known for many of the PNe in our sample. 
  
  Excitation class of a PN is an important parameter which represents the nebular spectral class. It is
 related to several parameters of the PN like the nebular structure, mass, chemical compositions and the
 temperature and luminosity of its central star as discussed by Gurzadyan \& Egikyan (1991). Using the
 relations given by them to derive the excitation classes of PNe, we attempt to see if there are differences
 in the distributions of spectral classes of [WR] PNe, $wels$-PNe and normal PNe. We have determined the
 excitation classes (from 1 to 3) using the line fluxes of [OIII] (at 4959 \AA~ and at 5007 \AA) and H$\beta$
 and higher excitation classes (from 4 to 12) were derived from the line fluxes of [OIII] and He II (at 4686 \AA). Though,
 the ionic emission line fluxes for most PNe were from Acker, Ochsenbein, Stenholm et al. (1992), for some
 PNe they are taken from other published results (see the references listed in Table 2). In Fig 8 we show
 the histograms of excitation classes for 54 [WR] PNe, 55 $wels$-PNe and 84 normal-PNe. A two sample
 Kolmogorov-Smirnov test was performed between the distributions of excitation classes for different groups
 of PNe to see if they are similar. The value of the test statistics between the distributions of [WR] PNe
 and $wels$-PNe is $D=0.067$ and the probability to find the test statistics below the critical value is
 $P=0.99$. This shows that the data sample of excitation classes of [WR] PNe and $wels$-PNe come from similar
 distribution. A Kolmogorov-Smirnov test performed between the distributions of [WR] PNe and normal-PNe shows
 $D=0.21, P=0.89$ and between $wels$-PNe and normal PNe shows the values of $D=0.267, P=0.70$. Hence, the
 excitation class distribution of normal PNe is different from that of the [WR] PNe and $wels$-PNe. The
 emission line flux ratios of [OIII] to H$\beta$ are similar (smaller than 15) for all PNe except A 30 
 (ratio = 40, has a $wels$ central star), A 78 (ratio = 33 in the inner knot; Jacoby 1983; has a $wels$
  central star) and IRAS 15154-5258 (ratio = 74 for the central region, Harrington 1996; has a [WR] central
 star). The large ratio of [OIII] to H$\beta$ fluxes implies that dust heating could be significant in the
 nebula (Harrington 1996). Another PN  which shows a large ratio (ratio = 102) is IRAS 18333-2357 (Gillett 
 et al. 1989; Muthumariappan, Parthasarathy \& Ita 2013), however, its  central star identification as a 
 H-poor star is uncertain. 

 \vspace{2.0cm} \protect \begin{figure*} 
 \includegraphics[width=\textwidth]{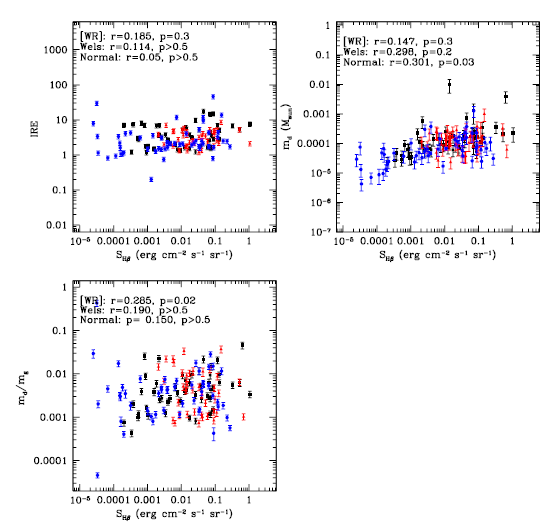}
 \caption{The dust parameters of [WR] PNe, $wels$-PNe and normal PNe plotted against $S_{H\beta}$. a) IRE
  (top left); b) $m_{dust}$ (top right); c) m$_{d}/m_{g}$ (bottom left). Data description as given in Fig. 1. See the text for details.} 
 \end{figure*} 

\subsection{Very small dust grains} 

The excess emission over the thermal equilibrium dust emission from a PN in the 1-15$\mu$m region should arise from very small grain population (VSG). These grains thermally fluctuate by absorbing UV photons 
from the hot CSPN and attain high temperatures ($\sim$ 1000 K) then they cool down by emitting at longer wavelengths. 
VSG emit significantly in the near-IR to the N band. The presence of VSG has been seen in  circumstellar environments, for example, in PNe A30 (Borkowski et al. 1994) and IRAS 18333-2357 (Muthumariappan, Parthasarathy \& Ita 2013). As seen in Section 3.1, larger fraction of [WR] PNe, in compared to other groups, are distributed close to the dust line in the near-IR CCDM ('ND' region in Fig 1) where the VSG contribute significantly. This hot dust component would come from the winds of the [WR] central stars (Gorny et al. 2001), similar to that was suggested for the near-IR colours of the population I WC stars (Williams 
et al. 1987). A good fraction of $wels$-PNe also show this hot dust emission. However, the average values of $WISE$ [3.4-4.6] colours for the three groups of PNe are very close, taking their dispersions into account (see Table 1).

\newpage
\begin{table} 
\label{tbl2}
\caption{Mean values of the parameters of [WR] PNe, $wels$-PNe and normal PNe along with their dispersion. The number of candidates were given in brackets.} 
\begin{tabular}{lrrr}
\hline
\hline
Parameter  & [WR] PNe & $wels$-PNe   & normal PNe  \\
(mean value) &  (No. of PNe)  & (No. of PNe)  &  (No. of PNe)  \\ 
\hline
           &         &         &        \\
 $T_{d}$(K)& 96$\pm$25 (78)  & 95$\pm$14 (52)& 93$\pm$31 (67)    \\ 
           & 75$\pm$13   & 75$\pm$9 & 74$\pm$17  \\ 
log[$m_{d}(M_{\odot})]$ & -3.83$\pm$0.52 (70) & -3.86$\pm$0.35 (49)& -4.33$\pm$0.56 (61) \\
                   & -3.57$\pm$0.51 & -3.59$\pm$0.35 & -4.06$\pm$0.56 \\
log[$m_{d}/m_{g}$] &  -2.38$\pm$0.63 (52) & -2.37$\pm$0.50 (43) & -2.51$\pm$0.58 (48) \\  
               &   -2.19$\pm$0.60 & -2.17$\pm$0.50 & -2.30$\pm$0.57  \\
 log[$n_{e}(cm^{-3}])$ & 3.54$\pm$0.46 (45) & 3.63$\pm$0.41 (29) & 3.52$\pm$0.54 (31)     \\
 log[$L_{IR}(L_{\odot})$] & 2.91$\pm$0.55 (67) & 2.90$\pm$0.32 (49) & 2.76$\pm$0.50 (61) \\
          &  2.88$\pm$0.53  & 2.88$\pm$0.32 & 2.72$\pm$0.47      \\          
 log[IRE] &  0.64$\pm$0.60 (54) & 0.47$\pm$0.22 (42)   & 0.43$\pm$0.40 (52) \\ 
          &  0.62$\pm$0.58 & 0.45$\pm$0.22 & 0.42$\pm$0.40  \\
          &         &         &          \\
\hline
\end{tabular}
\end{table}


 \vspace{2.0cm} \protect 
\begin{figure*}
 \includegraphics[width=\textwidth]{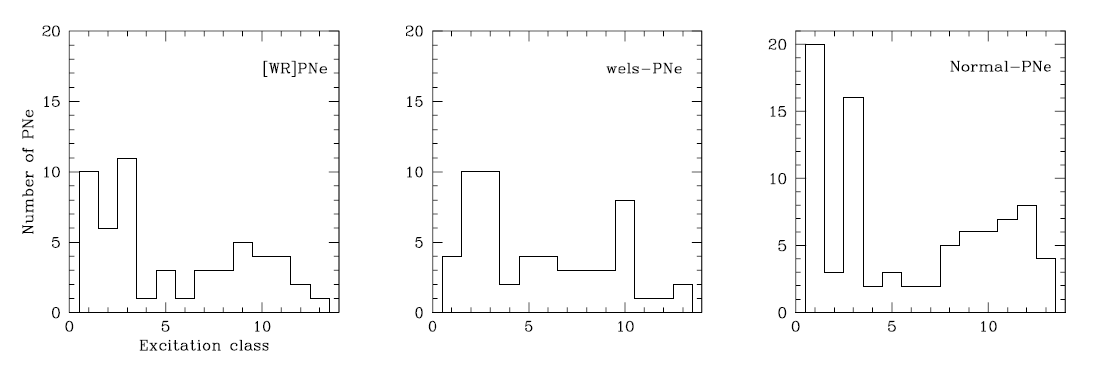}
 \caption{Histograms of excitation classes of [WR] PNe, $wels$-PNe and normal PNe. See the text  for details.} 
\end{figure*}


\newpage
\section{Discussion}
We discuss here using  our results and those available in the literature if there is a possibility for evolutionary sequences between different groups of PNe considered in this study. 
 
\subsection{Evolutionary Sequence between the subtypes of [WR] PNe}
  
 AN03 showed that the central star temperatures of [WR] PNe, derived from energy-balance method, increase along the 
 [WC] to [WO] evolutionary sequence. Gorny et al. (2000) plotted $S_{H\beta}$ of the subtypes of [WR] PNe 
 against their central star spectral types  which showed a steady decrease in nebular $S_{H\beta}$ 
 for PNe with central star types from [WCL] to [WCE]. Our dust colour temperature plot against $S_{H\beta}$ for 
 the subtypes of [WR] PNe in Fig 6b shows that the PNe with [WCL] central stars occupy the region with higher 
 values of $S_{H\beta}$ and $T_{d}$ while the PNe with [WCE] stars are seen in the region where they have lower values (though they show good overlap in the intermediate values)
 and the plot shows a continuity. 
 Further, there is no significant difference in the distribution of PNe with [WCL] and [WCE] central stars in the 'NS' and 
 'ND' regions of the near-IR CCDM shown in Fig 1b. The figure also shows that there is no [WCL] PN in the nebular box.
 This may imply that nebular emission becomes more significant as [WR]PNe evolve. Hence this study supports the proposal
 given earlier by AN03 and Girard, Koppen \& Acker (2007) that [WCL] and [WCE] are evolutionarily connected. However, it 
 should be noted that all the well studied PNe with [WCL] central stars show dual dust chemistry which is not commonly observed 
 in PNe with [WCE] central stars (De Marco 2008). 

\subsection{Is there a Evolutionary Sequence between [WR] PNe and $wels$-PNe ? }

 From their stellar terminal velocity $vs$ effective temperature diagram, AN03 suggested that the [WR] sample shows a 
 mean central star mass that is higher than the value shown by normal CSPNe. However, on the age-temperature diagram,
 Gesicki et al. (2006) found that $wels$ are located on tracks of both high and low stellar mass while [WR] stars take 
 a range of intermediate masses. They also find that nebular turbulence is almost universal for [WR] PNe and commonly 
 observed for $wels$-PNe but it is quite rare for normal PNe. They suggested that one group of $wels$ may have an evolutionary
 connection with the [WO] stars as the photospheric temperature increases, otherwise [WR] and $wels$ appear to represent
 independent evolutionary tracks. Girard, Koppen \& Acker (2007) found from their analysis of chemical  compositions of a
 sample of [WR] PNe and $wels$-PNe that $wels$-PNe evidently belong to a separate subclass of PNe, and unlikely they have
 evolutionary connection with [WR] PNe.\\
 
\noindent
 From Fig 6a  and Fig 6c we find that the distribution of dust colour temperatures and IR luminosities for
 [WR] PNe, $wels$-PNe and normal PNe against $S_{H\beta}$ are continuous for each group. In other words,
 removing one group of PNe do not make the plots discontinuous, suggesting that they don't
 form an evolutionary sequence. The Galactic latitudinal distribution and the excitation class distribution
 of [WR] PNe and $wels$-PNe are quite similar and they are distinctly different  from that of the normal PNe.
 This reaffirms that these three groups of PNe make three PNe populations and also indicates that [WR] PNe,
 $wels$-PNe do not form an evolutionarily sequence. The mean values of $m_{d}$ are similar for [WR] PNe and
 $wels$-PNe whereas normal-PNe have a tendency towards a lower mean value for mean $m_{d}$. These observed
 similarities show some underlying common features between [WR] PNe and $wels$-PNe, which are not owned by
 normal ones, which may be related to their origin. Hence, our study supports the proposal that [WR] PNe,
 $wels$-PNe and normal PNe are not evolutionarily connected but in some way [WR] PNe and $wels$-PNe are
 related. However, the 2MASS CCDM shows that the distribution of [WR] PNe is somewhat different from that 
 of $wels$-PNe. [WR] PNe commonly show strong IR emission features but $wels$-PNe show the weakest emission.
 The dispersions shown by $wels$-PNe in the mean values of parameters given in Table 1 and Table 5 are
 narrower than their respective dispersions shown by other two groups suggesting that they are relatively
 homogeneous.

 A handful of PNe show H-poor ejections which are located inside their (older) H-rich  outer shells. Among five such
 PNe, four of them have their central stars classified. Two PNe (A30 and A78) have $wels$ central stars and
 other two (A58 and IRAS 15154-5258), which are relatively younger, have [WR] central stars. The values of 
 $m_{d}$ and $L_{IR}$ of two of these H-poor PNe are quite high: they are 1.08$\times10^{-3}$M$_{\sun}$ and
 4780L$_{\sun}$ for A58 and 1.02$\times10^{-3}$M$_{\sun}$ and 2502L$_{\sun}$ for A30. $IRAS$ measurements 
 for A78 are not available and hence these values were not calculated. Zijlstra (2001) suggested an
 evolutionary sequence of A58 $\rightarrow$ IRAS 15154-5258 $\rightarrow$ A30/A78 and argued the weaker 
 winds in A30/A78 could be due to their declining luminosity. Though our study argues that [WR] PNe and 
 $wels$-PNe do not form an evolutionary sequence, it requires still more work to be done addressing all the
 related findings to reach a final conclusion. 
 
\section{Conclusions}

The main conclusions of the paper are the following:

\noindent
1) In the near-IR CCDM, [WR] PNe are usually seen away from the nebular box whereas, $wels$-PNe are commonly 
seen inside the nebular box. [WR] PNe have more tendency of finding them towards the dust line when compared to other PNe. \\ 

2) [WR] PNe show a tendency towards brighter $IRAS$ 12$\mu$m band in compared to $wels$-PNe and normal PNe, which possibly show that they have strong PAH emission. \\

3) Cool AGB dust in PNe traced by $IRAS$ 25- and 60$\mu$m bands have very similar colour for all the three
   groups of PNe. \\

4) All PNe show a tight correlation of decrease in $T_{d}$ with nebular age. $wels$-PNe are not seen at
   dust temperature lower than 77K whereas normal PNe have dust temperature as low as 59K. 
   Minimum H$\beta$ surface brightness seen for [WR] PNe and $wels$-PNe are significantly 
   larger than the value shown by normal-PNe, implying that they are relatively younger.
    While $L_{IR}$ of PNe are strongly correlated (decrease) with their age, $m_{d}$ and $m_{d}/m_{g}$ do 
   not change noticeably as the PN evolves. \\

5) [WR] PNe and $wels$-PNe are relatively distributed closer to the Galactic plane in compared to the normal PNe.
  The excitation class distributions of [WR] PNe is quite similar to that of $wels$-PNe and both have significant
  deviation from the distribution shown by normal PNe. \\

\section{Acknowledgements}

This research has made use of the SIMBAD database, operated at CDS, Strasbourg, France. 
This publication makes use of data products from the Two Micron All Sky Survey, which is a joint project of the
University of Massachusetts and the Infrared Processing and Analysis Centre, funded by the National Aeronautics 
and Space Administration and the National Science Foundation. This work is based in part on observations made 
with $WISE$, obtained from the NASA/IPAC Infrared Science Archive, operated by the Jet Propulsion Laboratory, 
California Institute of Technology, under contract with the National Aeronautics and Space Administration.
The authors thank the referee for his/her comments and suggestions.


\label{lastpage}

\end{document}